\documentclass[conf]{new-aiaa}
\usepackage[utf8]{inputenc}

\usepackage{graphicx}
\usepackage{amsmath}
\usepackage[version=4]{mhchem}
\usepackage{siunitx}
\usepackage{longtable,tabularx}
\usepackage{subcaption}
\usepackage{bbm}
\usepackage{comment}
\title{Wavelet-based resolvent analysis for statistically-stationary and temporally-evolving flows}
\author{Eric Ballouz \footnote{Ph.D. Student, Mechanical and Civil Engineering, AIAA Student Member}}
\affil{California Institute of Technology, Pasadena, CA, 91125}

\author{Barbara Lopez-Doriga \footnote{Ph.D. Student, Mechanical, Materials, and Aerospace Engineering Department, AIAA Student Member} and Scott T. M. Dawson \footnote{Assistant Professor, Mechanical, Materials, and Aerospace Engineering Department, AIAA Senior Member}}
\affil{Illinois Institute of Technology, Chicago, IL, 60616}

\author{H. Jane Bae \footnote{Assistant Professor, Graduate Aerospace Laboratories, AIAA Senior Member}}
\affil{California Institute of Technology, Pasadena, CA, 91125}

\begin{document}

\maketitle

\begin{abstract}
This work introduces a formulation of resolvent analysis that uses wavelet transforms rather than Fourier transforms in time. This allows resolvent analysis to be extended to turbulent flows with non-stationary means in addition to statistically-stationary flows. The optimal resolvent modes for this formulation correspond to the potentially time-transient structures that are most amplified by the linearized Navier-Stokes operator. We validate this methodology for turbulent channel flow and show that the wavelet-based and Fourier-based resolvent analyses are equivalent for statistically-stationary flows. We then apply the wavelet-based resolvent analysis to study the transient growth mechanism in the buffer layer of a turbulent channel flow by windowing the resolvent operator in time and frequency. The method is also applied to temporally-evolving parallel shear flows such as an oscillating boundary layer and three-dimensional channel flow, in which a lateral pressure gradient perturbs a fully-developed turbulent flow in a channel.
\end{abstract}

\section{Nomenclature}

{\renewcommand\arraystretch{1.0}
\noindent\begin{longtable*}{@{}l @{\quad=\quad} l@{}}
$x_1$ & streamwise direction\\
$x_2$ & wall-normal direction\\
$x_3$ & spanwise direction\\
$\bar{u}_i$ & total velocity in the $x_i$ direction\\
$U_i$ & average velocity over ensembles and homogeneous directions in the $x_i$ direction\\
$u_i$ & fluctuating velocity in the $x_i$ direction\\
$\bar{p}$ & total pressure\\
$P$ & average pressure over ensembles and homogeneous directions\\
$p$ & fluctuating pressure\\
$\nu$ & kinematic viscosity\\
$\rho$ & density\\
$\langle\cdot\rangle$ & average over ensembles and homogeneous directions \\
$F$, $(\hat{\cdot})$ & Fourier transform in homogeneous directions and time\\
$W$, $(\tilde{\cdot})$ & wavelet transform in time and Fourier transform in homogeneous directions\\
$L$ & discretized Laplacian\\
$D_t$ & discretized temporal derivative\\
$D_i$ & discretized spatial derivative in the $x_i$ direction\\
$T$ & time horizon\\
$N_{x_i}$ & number of spatial points in the $x_i$ direction\\
$\Delta x_i$ & grid size in the $x_i$ direction\\
$N_t$ &  number of discrete temporal points\\
$k_i$ & wavenumber in the $x_i$ direction \\
$\omega$ & temporal frequency \\
$\varphi$ & wavelet function \\
$\alpha$ & wavelet scale parameter \\
$\beta$ & wavelet shift parameter \\
$\hat{\mathcal{H}}, \tilde{\mathcal H}$ & resolvent operator (Fourier- and wavelet-based resolvent, respectively)\\
$ \hat \phi, \tilde \phi$ & principal resolvent forcing mode (for the Fourier- and wavelet-based resolvent, respectively)  \\
$\hat \psi, \tilde \psi$ & principal resolvent response mode (for the Fourier- and wavelet-based resolvent, respectively) \\
$\psi_i$ & $i$-th component of the principal resolvent response mode \\
$\sigma$ & principal resolvent singular value \\
$\mathcal{B}$ & windowing matrix that restrict the forcing to a subset of the full space\\
$\mathcal{C}$ & windowing matrix that restrict the response to a subset of the full space\\
$(\cdot)^\dagger$ & Moore-Penrose pseudo-inverse\\
$\delta$ & channel half-height \\
$Re_\tau$ & friction Reynolds number \\
$u_\tau$ & friction velocity \\
$(\cdot)^+$ & viscous units \\
$\Omega$ & channel wall oscillation frequency for the Stokes boundary layer\\
$Re_\Omega$ & Reynolds number for the Stokes boundary layer \\
$\delta_\Omega$ & Stokes laminar boundary layer thickness\\
$U_{\max}$ & maximum streamwise wall velocity for the Stokes boundary layer \\
$U_{i, rms}$ & root-mean-square velocities profiles in the $x_i$ direction \\
$\Pi$ & ratio of the mean spanwise and streamwise pressure gradient\\
$u_{\tau, 0}$ & friction velocity at time $t = 0$ \\
$u_{\tau, T}$ & friction velocity at time $t = T$ \\
$\tau_i$ & wall-shear stress in the $x_i$ direction\\
$\gamma$ & wall-shear angle $\tan^{-1}(\tau_3/\tau_1)$\\

\end{longtable*}}

\section{Introduction}

Fundamental studies of unsteady aerodynamics and turbulent flow have mostly focused on statistically stationary configurations, where all statistics are invariant under a shift in time. However, in real-world applications of external aerodynamics, truly unsteady transient effects become important. Such events include separation leading to stall, a sudden change in yaw angle, gust encounters, and shockwave formation. The additional complexity of these transient problems makes it harder to perform controlled experiments, both numerical and in the laboratory. Our goal is to develop reduced-order models that can be used to understand, predict, and control highly-unsteady transient turbulent flow in various engineering systems.

Resolvent analysis has been a popular reduced-order model for understanding a wide variety of turbulent flows. Resolvent analysis refers to the inspection of the resolvent operator, a linear operator that consists of the linearized Navier-Stokes and maps forcing inputs (which can be due to the nonlinear advection terms, or other exogenous inputs) onto the flow states. This operator governs how inputs are amplified by the linear dynamics of the system. Its singular value decomposition (SVD) identifies the inputs to which the linearized equations of motion are most receptive, their gains, and the most amplified outputs \citep{Jovanovic2005,McKeon2010}. Often, the truncated SVD is enough to capture most of the effect of the operator due to the fast decay of the singular values. The resulting low-rank approximation of the forcing-response dynamics of the full system is extremely valuable for modeling, controlling, and understanding the physics of fluid flows \citep{sharma2013coherent,moarref2013model,luhar2014opposition,martini2020resolvent,towne2020resolvent,bae2021nonlinear,yeh2019resolvent}, and has been extended to compressible and stratified flows \citep{bae2020resolvent,bae2020studying,ahmed2021resolvent}. However, a Fourier transform in time is traditionally used to formulate the resolvent operator, which restricts its analysis and application to statistically-steady and quasi-periodic flows \citep{Padovan2020}. Indeed, the resulting SVD modes will be Fourier modes in time, and cannot represent temporally local effects. 

To construct the resolvent operator, we instead propose using a wavelet transform \citep{Meyer1992} in time. The most effective inputs and the most amplified outputs given by the SVD of the newly-formulated resolvent operator will be spatiotemporally localized.
There have been numerous research efforts applying wavelet methods in fluid mechanics, with early work predating the application of resolvent analysis to turbulent flows and focusing on spatial transformations \cite{meneveau1991analysis,lewalle1993wavelet}. More recent work has focused on data-driven wavelet transforms \cite{ren2021image,floryan2021discovering}. The proposed research effort will provide an important bridge connecting such data-driven analyses to the underlying physics, analogous to recent work connecting and comparing resolvent analysis to data-driven spectral proper orthogonal decomposition \cite{towne2018spectral,abreu2020spectral,tissot2021stochastic}.

In the present work, we develop and validate wavelet-based resolvent analysis for a variety of systems, ranging from quasi-parallel wall-bounded turbulent flows to spatio-temporally evolving systems. In \S\ref{sec:formulation}, we formulate the Fourier- and wavelet-based resolvents; we highlight the properties of the wavelet transform and justify the choice of wavelet basis. In \S\ref{sec:application} we first establish the equivalence of Fourier- and wavelet-based resolvent analyses for the statistically stationary turbulent channel flow and showcase the additional capacity of the wavelet-based resolvent to capture the linear transient growth of streaks under transient forcing. We also apply wavelet-based resolvent analysis to statistically non-stationary flows in \S\ref{sec:app_time}, notably the Stokes boundary layer flow as well as a turbulent channel flow subjected to a sudden lateral pressure gradient. Finally, the summary of the work and a discussion of the results are given in \S\ref{sec:conclusion}.

\section{Mathematical formulation} \label{sec:formulation}

\subsection{Fourier-based resolvent analysis}

The incompressible Navier-Stokes equations in Einstein notation are given by 
\begin{equation}
    \frac{\partial\bar{u}_i}{\partial{t}} + \bar{u}_j\frac{\partial\bar{u}_i}{\partial{x_j}} = -\frac{1}{\rho}\frac{\partial\bar{p}}{\partial x_i} + \nu\frac{\partial^2\bar{u}_i}{\partial{x}_j\partial{x}_j},\quad \frac{\partial\bar{u}_i}{\partial{x_i}} = 0,
    \label{eq:NS}
\end{equation}
where $\bar{u}_i$ is the total velocity (including the mean and the fluctuating component), $\tilde{p}$ is the total pressure, $\rho$ is the density, and $\nu$ is the kinematic viscosity. The total velocity can be divided into $\bar{u}_i = U_i + u_i$, where $U_i$ is the average over ensembles and homogeneous directions, and $u_i$ is the fluctuating component. Similarly, pressure can be decomposed as $\tilde{p} = P + p$.

Equation~(\ref{eq:NS}) can then be divided into the mean and the fluctuating equations
\begin{gather}
    \frac{\partial{U_i}}{\partial t} + \left\langle \bar{u}_j\frac{\partial\bar{u}_i}{\partial{x_j}}\right\rangle = -\frac{1}{\rho}\frac{\partial{P}}{\partial{x_i}} + \nu \frac{\partial^2U_i}{\partial{x_j}\partial{x_j}},\quad \frac{\partial{U_i}}{\partial{x_i}}= 0,\\
    \frac{\partial{u_i}}{\partial t} + U_j\frac{\partial{u_i}}{\partial{x_j}} + u_j\frac{\partial{U_i}}{\partial{x_j}} = -\frac{1}{\rho}\frac{\partial{p}}{\partial{x_i}} + \nu \frac{\partial^2u_i}{\partial{x_j}\partial{x_j}} + f_i,\quad \frac{\partial{u_i}}{\partial{x_i}}= 0,
\end{gather}
where $f_i$ is the remaining nonlinear terms in the fluctuating equations and $\langle \cdot \rangle$ is the average over ensembles and in homogeneous directions. Note that some of the terms in the fluctuating equations may be zero depending on the flow configuration. The equations above do not have an analytic solution unless in very particular situations and are most commonly solved numerically. Discretizing the fluctuating equations, we get
\begin{equation}
    D_t{u_i} + U_jD_j{u_i} + u_jdU_{i,j} = -\frac{1}{\rho}D_i{p} + \nu L{u_i} + f_i,\quad D_iu_i = 0,
    \label{eq:Disc_NS}
\end{equation}
where $D_t$ is the discrete derivative in time, $D_i$ is the discrete derivative in the $x_i$ direction, $L$ is the discrete Laplacian, and $dU_{i,j}$ denotes the matrix form of $\partial U_i/\partial x_j$. Each discretized equation is an $N_t \times N_{x_1} \times N_{x_2} \times N_{x_3}$-dimensional system, where $N_t$ is the temporal resolution, and $N_{x_i}$ are the spatial resolutions in the $x_i$, $i=1,2,3$ directions respectively. The discretized velocity and velocity gradient, $U_j$ and $dU_{i, j}$, are $(N_t \times N_{x_1} \times N_{x_2} \times N_{x_3})^2$ diagonal matrices. In the traditional resolvent analysis, we apply the Fourier transform operator in homogeneous directions and time, $F$, to the left in Eq.~(\ref{eq:Disc_NS}), leading to the equation
\begin{multline}
    (FD_tF^{-1})(F{u_i}) + (FU_jD_jF^{-1})(F{u_i}) + (Fu_j)(FdU_{i,j}F^{-1}) =\\
    -\frac{1}{\rho}(FD_iF^{-1})(F{p}) + \nu (FLF^{-1})(F{u_i}) + Ff_i,\quad (FD_iF^{-1})(Fu_i) = 0,
\end{multline}
where $F^{-1}$ is the inverse transformation, or equivalently 
\begin{equation}
    \widehat{D_t}\hat{u}_i + \widehat{U_jD_j}\hat{u}_i + \widehat{dU_{i,j}}\hat{u}_j =
    -\frac{1}{\rho}\widehat{D_i}\hat{p} + \nu \hat{L}\hat{u}_i + \hat{f}_i,\quad \widehat{D_i}\hat{u}_i = 0.
\end{equation}
Note that for an arbitrary matrix $M$ and vector $b$, $\hat{M} := F M F^{-1}$ and $\hat{b}:= Fb$. For temporally stationary systems, this equation can typically be decoupled for each wavenumber and frequency combination. For example, in the case of channel flow where the flow is homogeneous in the streamwise ($x_1$) and spanwise ($x_3$) directions, the linear operator can be cast as 
\begin{equation}
\left[\begin{array}{c}\hat{u}_1(k_1,k_3,\omega)\\ \hat{u}_2(k_1,k_3,\omega)\\ \hat{u}_3(k_1,k_3,\omega)\\ \hat{p}(k_1,k_3,\omega)\end{array}\right] = \hat{\mathcal{H}}(k_1,k_3,\omega) \left[\begin{array}{c}\hat{f}_1(k_1,k_3,\omega)\\ \hat{f}_2(k_1,k_3,\omega)\\ \hat{f}_3(k_1,k_3,\omega)\\ 0 \end{array}\right]
\label{eq:in_out_fourier}
\end{equation}
for a given $(k_1,k_3,\omega)$ triplet. 
Typically, the singular value decomposition of the linear operator $\hat{\mathcal{H}}\in \mathbb{C}^{4N_{x_2}} \times \mathbb{C}^{4N_{x_2}}$ is taken to study the left and right singular vectors as response and forcing modes, and the singular values as amplification factors or gains. 

\subsection{Wavelet-based resolvent analysis}

\subsubsection{Formulation}

To account for transient behavior in the mean flow or the fluctuations, we introduce the wavelet-based resolvent analysis. The benefit of the wavelet transform in time is that it preserves both time and frequency information. The wavelet transform projects a function onto a wavelet basis composed of scaled and shifted versions of a mother function $\varphi(t)$. The transformed function depends on the scale ($\alpha$) and shift ($\beta$) parameters respectively linked to frequency and time information, whereas the Fourier transform is a function of only frequency.

We propose using a wavelet transform in time while keeping the Fourier transform in homogeneous directions. We denote the total transformation operator (wavelet in time and Fourier in homogeneous directions) as $W$ and the left inverse operator as $W^{-1}$, which is also the right inverse for unitary transforms. The inverse operator is well-defined and unique for orthogonal wavelet bases. We can then apply $W$ on the left in Eq.~(\ref{eq:Disc_NS}), which gives
\begin{multline}
    (WD_tW^{-1})(W{u_i}) + (WU_jD_jW^{-1})(W{u_i}) + (WdU_{i,j}W^{-1})(Wu_j) =\\
    -\frac{1}{\rho}(WD_iW^{-1})(W{p}) + \nu (WLW^{-1})(W{u_i}) + Wf_i,\quad (WD_iW^{-1})(Wu_i) = 0,
\end{multline}
or
\begin{equation}
    \widetilde{D_t}\tilde{u}_i + \widetilde{U_jD_j}\tilde{u}_i + \widetilde{dU_{i,j}}\tilde{u}_j = -\frac{1}{\rho}\widetilde{D_i}\tilde{p} + \nu \tilde{L}\tilde{u}_i + \tilde{f}_i,\quad \widetilde{D_i}\tilde{u}_i = 0.
\end{equation}
Note that for an arbitrary matrix $M$ and vector $b$, $\tilde{M} := W M W^{-1}$ and $\tilde{b}:= Wb$. These equations can be separated for each spatial wavenumber in the homogeneous direction, and thus the dimension of each linear equation is smaller than the full Navier-Stokes equations. Note that if the transformation in time is given by the Fourier transform rather than a wavelet transform, this would recover the traditional Fourier-based resolvent analysis \citep{McKeon2010} (if the flow is temporally stationary) or the harmonic resolvent \citep{Padovan2020} analysis (if the flow is periodic in time). Similar to the Fourier-based resolvent analysis, for flows that are homogeneous in the $x_1$ and $x_3$ directions, this can be written in matrix form as
\begin{equation}
\left[\begin{array}{c}\tilde{u}_1(k_1,k_3)\\ \tilde{u}_2(k_1,k_3)\\ \tilde{u}_3(k_1,k_3)\\ \tilde{p}(k_1,k_3)\end{array}\right] = \tilde{\mathcal{H}}(k_1,k_3) \left[\begin{array}{c}\tilde{f}_1(k_1,k_3)\\ \tilde{f}_2(k_1,k_3)\\ \tilde{f}_3(k_1,k_3)\\ 0 \end{array}\right],
\label{eq:in_out_w}
\end{equation}
where the wavelet-based resolvent operator $\tilde{\mathcal{H}} \in \mathbb{C}^{4N_t\times N_{x_2}} \times \mathbb{C}^{4N_t\times N_{x_2}}$ is defined as
\begin{equation}
    \tilde{\mathcal{H}} = \left[\left(\tilde{D}_t - \nu \tilde{L} + \widetilde{U_iD_i}\right) 
    \left(
    \begin{array}{cccc}
    1 & 0 & 0 & 0 \\
    0 & 1 & 0 & 0 \\
    0 & 0 & 1 & 0 \\
    0 & 0 & 0 & 0 
    \end{array}
    \right)\right.
    +
    \left.\left(
    \begin{array}{cccc}
    \widetilde{dU}_{1,1} & \widetilde{dU}_{1,2} & \widetilde{dU}_{1,3} & \frac{1}{\rho}\tilde{D}_1 \\
    \widetilde{dU}_{2,1} & \widetilde{dU}_{2,2} & \widetilde{dU}_{2,3} & \frac{1}{\rho}\tilde{D}_2 \\
    \widetilde{dU}_{3,1} & \widetilde{dU}_{3,2} & \widetilde{dU}_{3,3} & \frac{1}{\rho}\tilde{D}_3 \\
    \tilde{D}_1        & \tilde{D}_2        & \tilde{D_3}        & 0 
    \end{array}
    \right)\right]^{-1}.
\end{equation}
This formulation allows us to study transient flows using resolvent analysis.

\subsubsection{Wavelet-based resolvent analysis with windowing} \label{sec:windowing}

We can reformulate a resolvent map between forcing and response at specific time shifts and scales by defining a windowed resolvent operator
\begin{equation}
\left[\begin{array}{c}\tilde{u}_1(k_1,k_3)\\ \tilde{u}_2(k_1,k_3)\\ \tilde{u}_3(k_1,k_3)\\ \tilde{p}(k_1,k_3)\end{array}\right] = \mathcal{C}\tilde{\mathcal{H}}(k_1,k_3) \mathcal{B} \left[\begin{array}{c}\tilde{f}_1(k_1,k_3)\\ \tilde{f}_2(k_1,k_3)\\ \tilde{f}_3(k_1,k_3)\\ 0 \end{array}\right],
\label{eq:windowed_res}
\end{equation}
where $\mathcal{B}$ and $\mathcal{C}$ are windowing matrices on the forcing and response modes, respectively \cite{jeun2016, kojima2020}. The windowing matrices select a subset of the full forcing and response states. For example, to select a particular scale and shift parameter $(\alpha_s,\beta_s)$ for the forcing mode, we set
\begin{equation}
    \mathcal{B} = \text{diag}\big(\mathbbm{1}(\alpha = \alpha_s)\mathbbm{1}(\beta = \beta_s)\big),
\end{equation}
where $\mathbbm{1}(\cdot)$ is an indicator function. The SVD of the windowed resolvent operator, $\mathcal{C}\tilde{\mathcal{H}}(k_1,k_3) \mathcal{B}$, allows us to identify forcing and response modes restricted to a limited frequency and time interval. 

The windowed resolvent operator, however, is equivalent to taking the Moore-Penrose pseudo-inverse of $\mathcal{B}^\dagger\tilde{\mathcal{H}}(k_1,k_3)^{-1}\mathcal{C}^\dagger$, where the superscript $\dagger$ indicates the pseudo-inverse. In this case, $\mathcal{B}^\dagger = \mathcal{B}$ and $\mathcal{C}^\dagger = \mathcal{C}$. By applying the pseudo-inverse to the linearized Navier-Stokes operator prior to inverting it to compute $\tilde{\mathcal{H}}(k_1,k_3)$, we can reduce the computational cost. The matrix inversion and SVD will apply to a matrix with its size defined by the nonzero block of $\mathcal{B}\mathcal{C}$ rather than the full system.

\subsubsection{Choice of wavelet basis}

\begin{figure}
\centering
\begin{subfigure}{0.45\textwidth}
    \includegraphics[width=\linewidth]{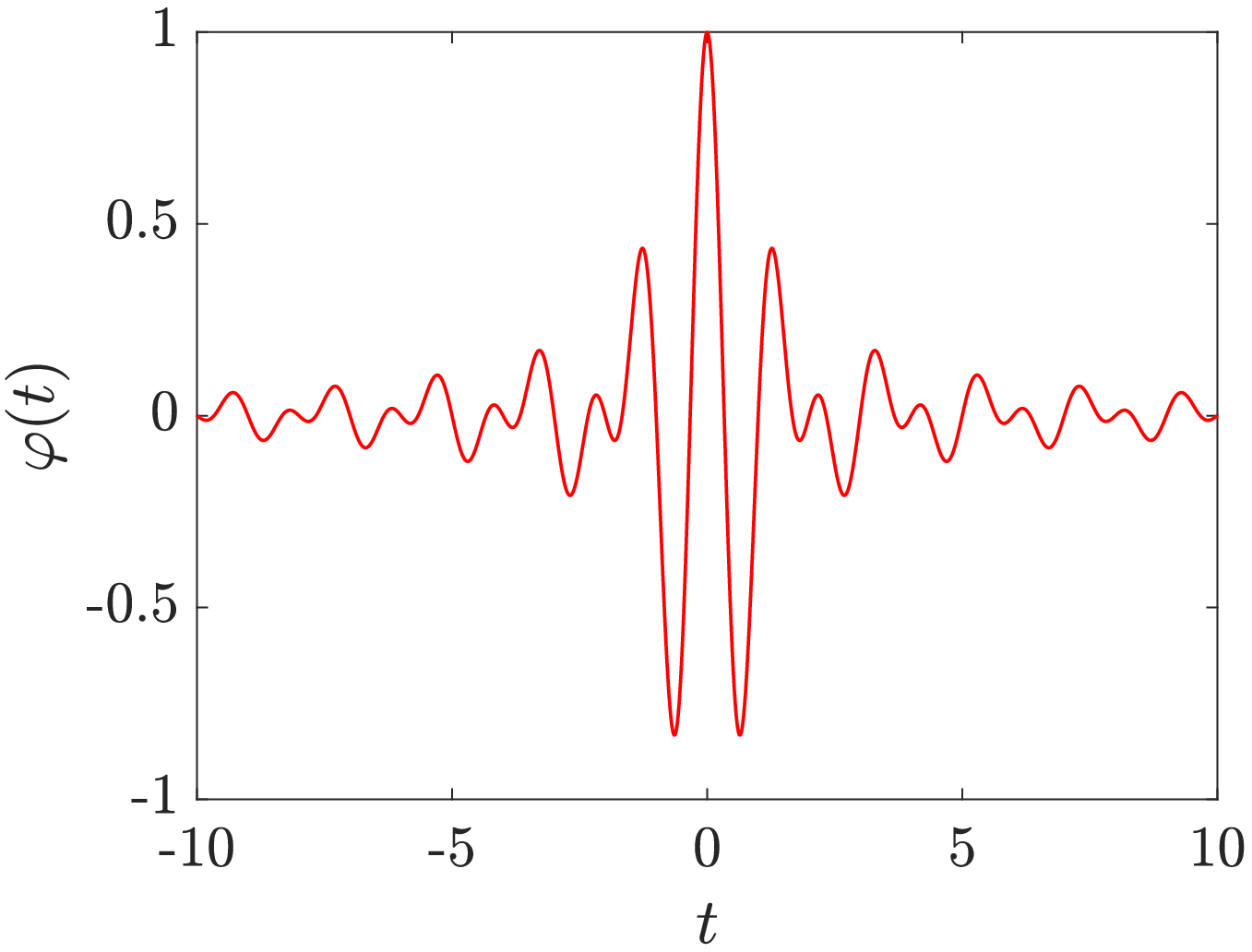}
    \caption{}
\end{subfigure}
\hspace{0.2cm}
\begin{subfigure}{0.45\textwidth}
    \includegraphics[width=\linewidth]{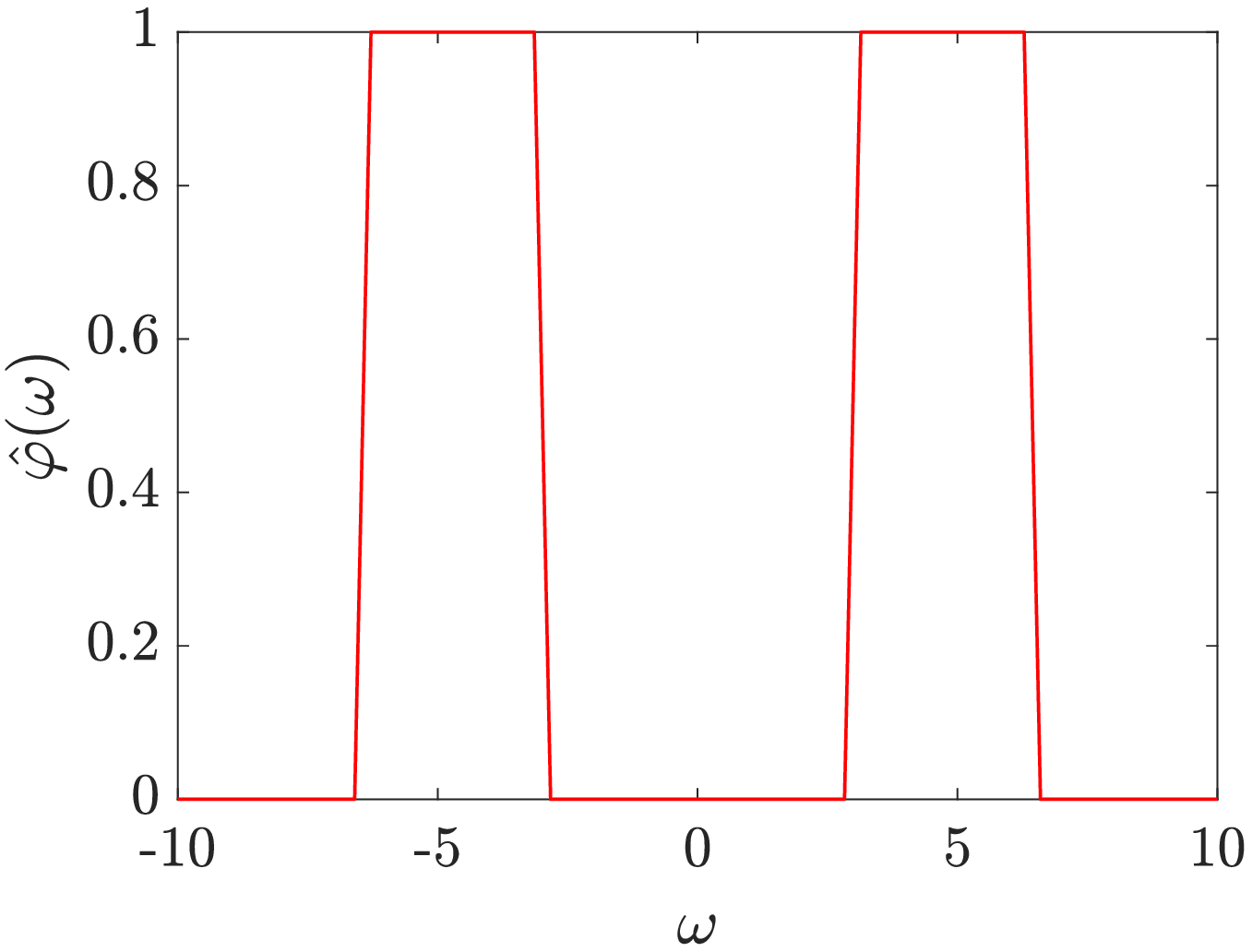}
    \caption{}
\end{subfigure}
\caption{Shannon wavelet in (a) time and (b) frequency domains.}
\label{fig:shannon}
\end{figure}

Wavelet transforms are not unique and are defined by the choice of the mother wavelet $\varphi(t)$. Consider an arbitrary function $f$ in $L_2(\mathbb R)$. Its Fourier and wavelet transforms are
\begin{equation}
\hat f(\omega) = \int_{-\infty}^{+\infty}f(t)e^{- \mathrm{i} \omega t} dt,
\end{equation}
\begin{equation}
\tilde f(\alpha, \beta) = \int_{-\infty}^{+\infty}f(t)\varphi^*_{\alpha, \beta}(t) dt,\quad\varphi_{\alpha,\beta}(t) \equiv \frac{1}{\sqrt{\alpha}}\varphi \left(\frac{t-\beta}{\alpha} \right).
\end{equation}
Note that there is a trade-off between precision in frequency and precision in time, i.e., one cannot find a function $\varphi(t)$ that is well localized in both time and frequency \cite{Mallat2001}. As two extreme examples, consider the Dirac delta centered at $t=1$, which is perfectly localized in space but with an infinite spread in frequency space, and the Fourier mode $e^{\mathrm{i}t}$, $\mathrm{i} = \sqrt{-1}$, which is perfectly localized in frequency space at $\omega = 1$ but has infinite spread in time.

For this study, we choose the Shannon wavelet for its frequency properties. Each Shannon wavelet acts as a perfect band-pass filter and covers a band $2^{-i}/T([-2\pi, -\pi] \cup [\pi, 2\pi])$, $i \in \mathbb{Z}$ (Fig. \ref{fig:shannon}), where $T$ denotes the time horizon considered. Though the Shannon wavelet does not have perfect frequency localization provided by the Fourier transform, it allows the separation of the frequency content into distinct non-overlapping bands for different scales. In addition, the discrete transform matrix $W$ \cite{Najmi2012, Mallat2001} for the Shannon wavelet is an orthonormal basis for $L_2(\mathbb R)$, making $W$ unitary \cite{Najmi2012}. However, we note that the results in the current study do not change significantly with different choices of the wavelet basis.

\subsubsection{Computational cost}

The construction of $\tilde{\mathcal{H}}$ requires the inversion of a $4N_y N_t \times 4N_y N_t$ matrix, a computation that costs $O(64 N_y^3 N_t^3)$ operations. The full SVD of $\tilde{\mathcal{H}}$ would also require $O(64 N_y^3 N_t^3)$ operations. With a direct solve, the wavelet-based resolvent analysis would cost $O(N_t^2)$-times more than performing $N_t$ separate Fourier-based resolvent for each temporal scale, though the latter would fail to capture the interactions between the different time scales. This penalty of $O(N_t^2)$ is the nominal cost of constructing time-localized resolvent modes. 

One method for reducing the memory and computational cost of wavelet-based resolvent analysis is to use sparse finite difference operators and wavelet transforms when constructing $\tilde{\mathcal{H}}^{-1}$. The resulting sparse matrix can be factored with specialized packages like PARDISO \cite{pardiso1, pardiso2, pardiso3}, which we utilize in the current study, in order to efficiently solve linear equations of the form $\tilde{\mathcal{H}}^{-1} v = w$, where $v$ and $w$ are arbitrary vectors. To further take advantage of the sparsity of $\tilde{\mathcal{H}}$, we opt for an iterative method to perform the SVD. In this work, we use a one-sided Lanczos bidiagonalization \cite{simon2000low}, an iterative algorithm that allows us to compute a truncated SVD and accurately estimate a number $q < 4N_yN_t$ of the most significant singular input and output modes. Other efficient SVD algorithms rely on randomized approaches, in particular by sub-sampling the high-dimensional matrix and performing the SVD on the lower-dimensional approximation \cite{halkotropp, drineas2016randnla, tropp2017}. A randomized SVD of a high-dimensional discrete resolvent operator was used in \cite{ribeiro2020randomized, yeh2020resolvent}.

In some cases such as \S\ref{sec:windowing}, where only some wavelet scales or time shifts are relevant, the windowing matrices $\mathcal C$ and $\mathcal B$ can be chosen to select the significant regions of the time-frequency domain. We then neglect the zero rows and columns in $\mathcal C \tilde{\mathcal H} \mathcal B$ associated with non-significant wavelet scales and shifts and perform the SVD and pseudo-inversion on the lower-dimensional system.

\section{Application to statistically-stationary flow} \label{sec:application}

We first validate the wavelet-based resolvent analysis on a statistically-stationary turbulent channel flow. We have shown that for unitary wavelet transforms the Fourier-based and wavelet-based resolvent modes are equivalent. Thus, for the channel flow case, we expect the two methods to produce identical resolvent modes. We then utilize the temporally-local property of wavelet-based resolvent analysis to study transient growth in a turbulent channel flow.

\subsection{Turbulent channel flow} \label{sec:app:channel}

\begin{figure}
\centering
\begin{subfigure}{0.45\textwidth}
    \centering
    \includegraphics[width=\linewidth]{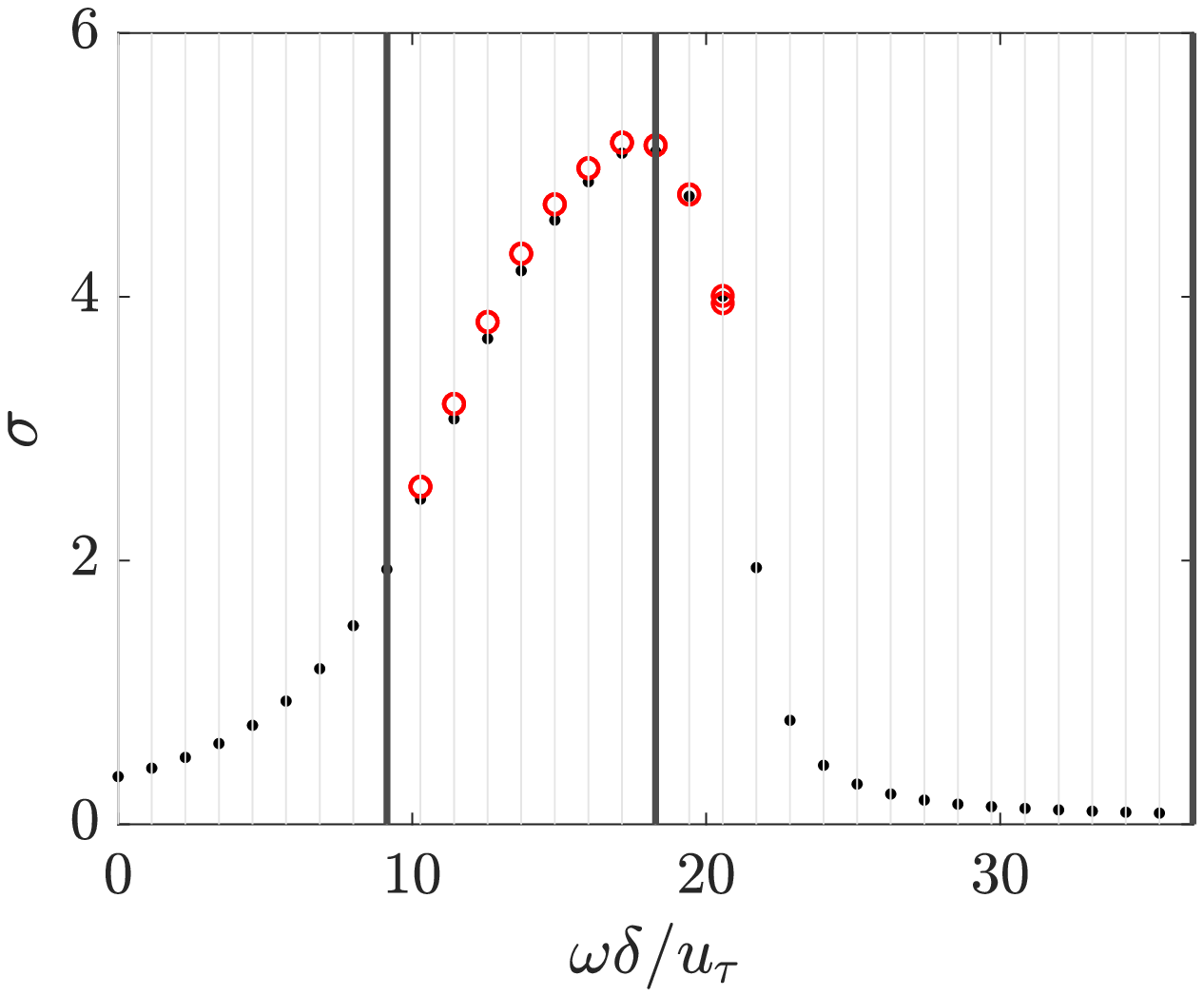}
    \caption{}
\end{subfigure} %
\hspace{0.2cm}
\begin{subfigure}{0.48\textwidth}
    \centering
    \includegraphics[width=\linewidth]{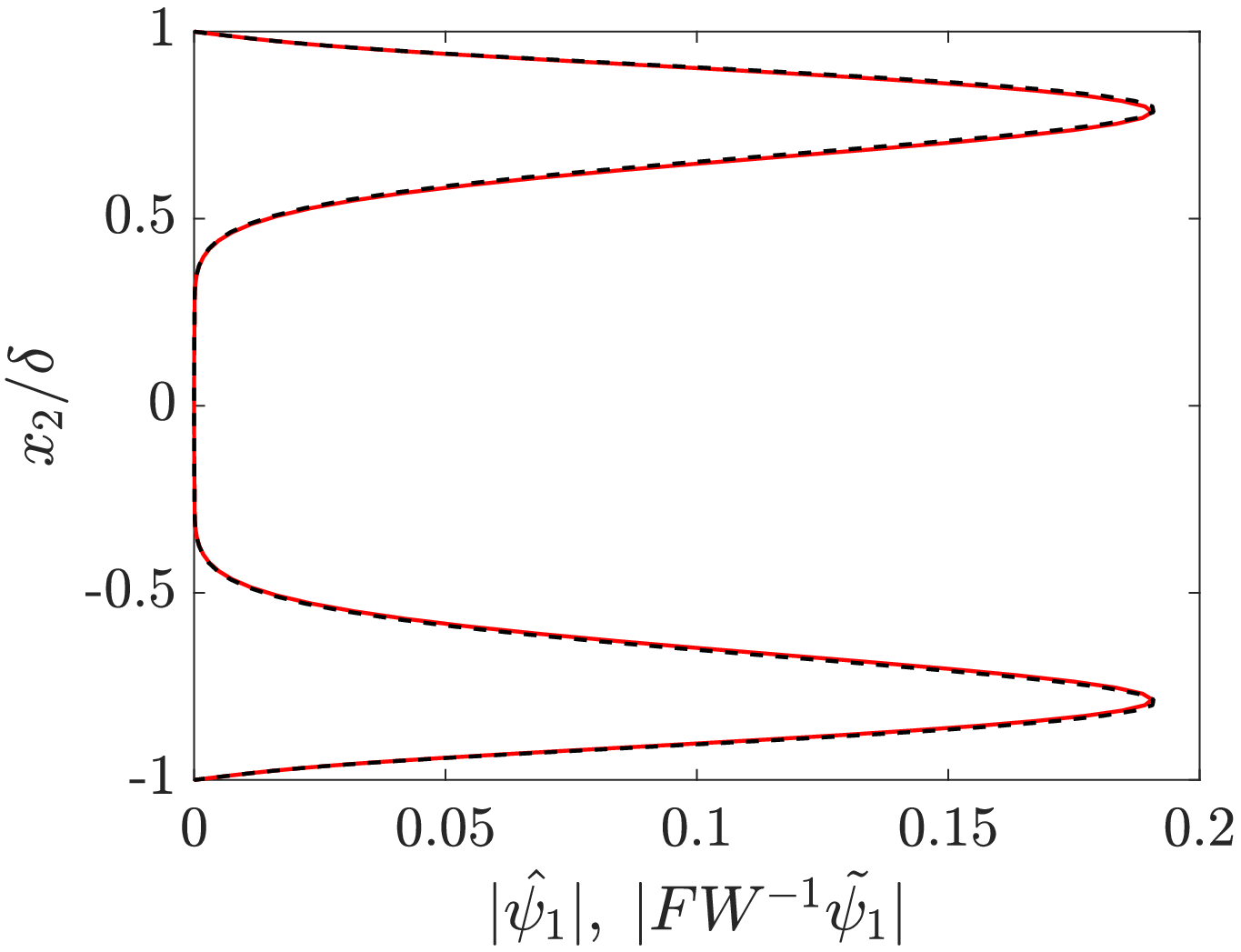}     
    \caption{}
\end{subfigure}

\caption{\label{fig:compare}(a) First ten singular values for the wavelet-based resolvent (red) and the largest singular value for the Fourier-based resolvent (black) computed for each $\omega_i$. The vertical gray lines indicate the temporal resolution in frequency space, and the vertical black lines delimit the frequency band covered by each of the chosen wavelet scales. (b) Magnitude of the Fourier-transformed wavelet-based (red) and Fourier-based (black) streamwise resolvent response mode for $\omega = 17.14 u_\tau/\delta$.}
\end{figure}

We first compare the results of both the traditional Fourier-based resolvent analysis and wavelet-based resolvent analysis. For this, we consider a turbulent channel flow. The mean profile of a turbulent channel flow at friction Reynolds number $Re_\tau = 186$ is obtained by a direct numerical simulation (DNS) using a second-order staggered finite-difference \cite{orlandi2000} and a fractional-step method \cite{kim1985} with a third-order Runge-Kutta time-advancing scheme \cite{wray1990}. Periodic boundary conditions are imposed in the streamwise and spanwise directions and the no-slip and no-penetration boundary conditions are used at the top and bottom walls. The code has been validated in previous studies in turbulent channel flows \cite{bae2018,bae2019,lozano-duran2019} and flat-plate boundary layers \cite{lozano-duran2018}. The numerical domain is $8\pi\delta \times 2\delta \times  3\pi\delta$ and is discretized using 768, 130, and 288 grid points in the streamwise, wall-normal, and spanwise directions, respectively. We use uniform grid spacing in the streamwise and spanwise directions of $\Delta x_1^+ = 6$ and $\Delta x_3^+ = 6$, and a wall-normal grid stretched away from the wall using a hyperbolic tangent with $\min(\Delta x_2^+) = 0.16$ and $\max(\Delta x_2^+) = 7.2$. Here the superscript $+$ denotes wall units defined in terms of $\nu$ and $u_\tau$, where $u_\tau$ is the friction velocity. The simulations were run for 100 eddy turnover times (defined as $\delta/u_\tau$) after transients to compute the mean quantities. 

For the resolvent analysis, the wall-normal direction is discretized using a Chebyshev collocation method using $N_{x_2} = 128$, and the mean streamwise velocity profile and its wall-normal derivative from the DNS are interpolated to the Chebyshev collocation points. We use a periodic boundary condition for the temporal domain, $T$, with a temporal resolution of $N_t = 128$. We choose $D_t$ to be a Fourier differentiation matrix, and $D_{x_2}$ a Chebyshev differentiation matrix with a no-slip and no-penetration boundary condition at the wall. Note that these matrices are not sparse; for this case, sparse differentiation matrices were not needed, though finite difference matrices may be used in higher-dimensional problems to improve efficiency. We choose the spanwise and streamwise wavelengths of $\lambda_1^+ \approx 1000$ and $\lambda_3^+ \approx 100$, in line with the most energetic structures close to the wall.

Since the mean profiles are statistically steady, we have $\widetilde{U_j D_j} = \widehat{U_j D_j}$, $\widetilde{dU_{i,j}} = \widehat{dU_{i,j}}$,  $\tilde L = \hat L$. The wavelet- and Fourier-based cases thus only differ by the time differentiation matrix such that $W^{-1}\widetilde{D_t}W = F^{-1}\hat D_t F = D_t$. Note that $W$ is unitary for the choice of wavelet. Since the singular value decomposition is unique up to multiplication by a unitary matrix, we expect the singular values of $\tilde{\mathcal H}$ to be the same as that of 
\begin{equation}
\hat{\mathcal{H}}(k_1,k_3) = 
\left (
\begin{array}{cccc}
\hat{\mathcal H}(k_1, k_3, \omega_1) & & & \\
& \hat{\mathcal H}(k_1, k_3, \omega_2) & & \\
& & \hat{\mathcal H}(k_1, k_3, \omega_3) & \\
& & & \ddots \\
\end{array} \right ),
\end{equation}
where $\omega_i = (2\pi i)/T$ for $i = -{N_t}/{2}, \cdots, {N_t}/{2}-1$. Here, we set $T = 5.5\delta/u_\tau$. Moreover, we expect the response and forcing modes of both systems to be related by the unitary transform given by the Fourier and inverse-wavelet transform in time, $FW^{-1}$. In Fig.~\ref{fig:compare}(a), we show the singular values of the leading Fourier- and wavelet-based resolvent response modes, $\hat \psi$ and $\tilde \psi$ respectively. The Fourier-based resolvent modes were computed by applying the resolvent analysis at each $\omega_i$ and taking the principal singular value. The wavelet-based resolvent modes were computed by solving the full space-time system at once. We also plot the most amplified streamwise resolvent response mode for the two methods in Fig.~\ref{fig:compare}(b). As expected, the singular values and the corresponding modes are equivalent. The small deviations in the singular values for the two cases are due to the truncation errors of the Shannon wavelet basis. 

\subsection{Transient growth mechanism of turbulent channel flow} \label{sec:transient}

\begin{figure}
    \centering
    \includegraphics[width=0.45\textwidth]{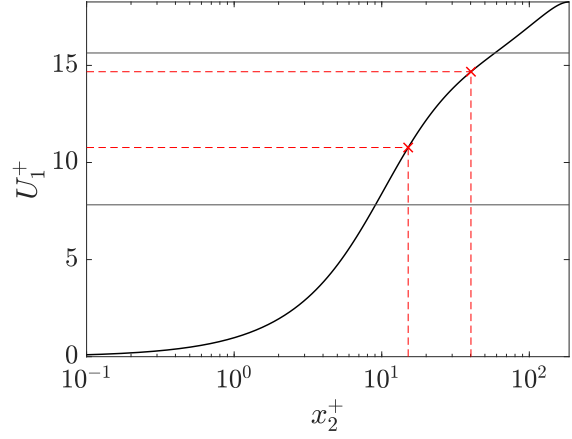}
    \caption{Mean streamwise velocity profile for channel flow at $Re_\tau = 186$. The buffer layer is delimited by the dashed lines; the black horizontal lines correspond to frequency bands covered by the chosen wavelets, mapped to streamwise velocities using $U_1 = \omega/k_1$.}
    \label{fig:buffer}
\end{figure}

\begin{figure}
\centering 
\begin{subfigure}{0.46\textwidth}
    \includegraphics[width=\linewidth]{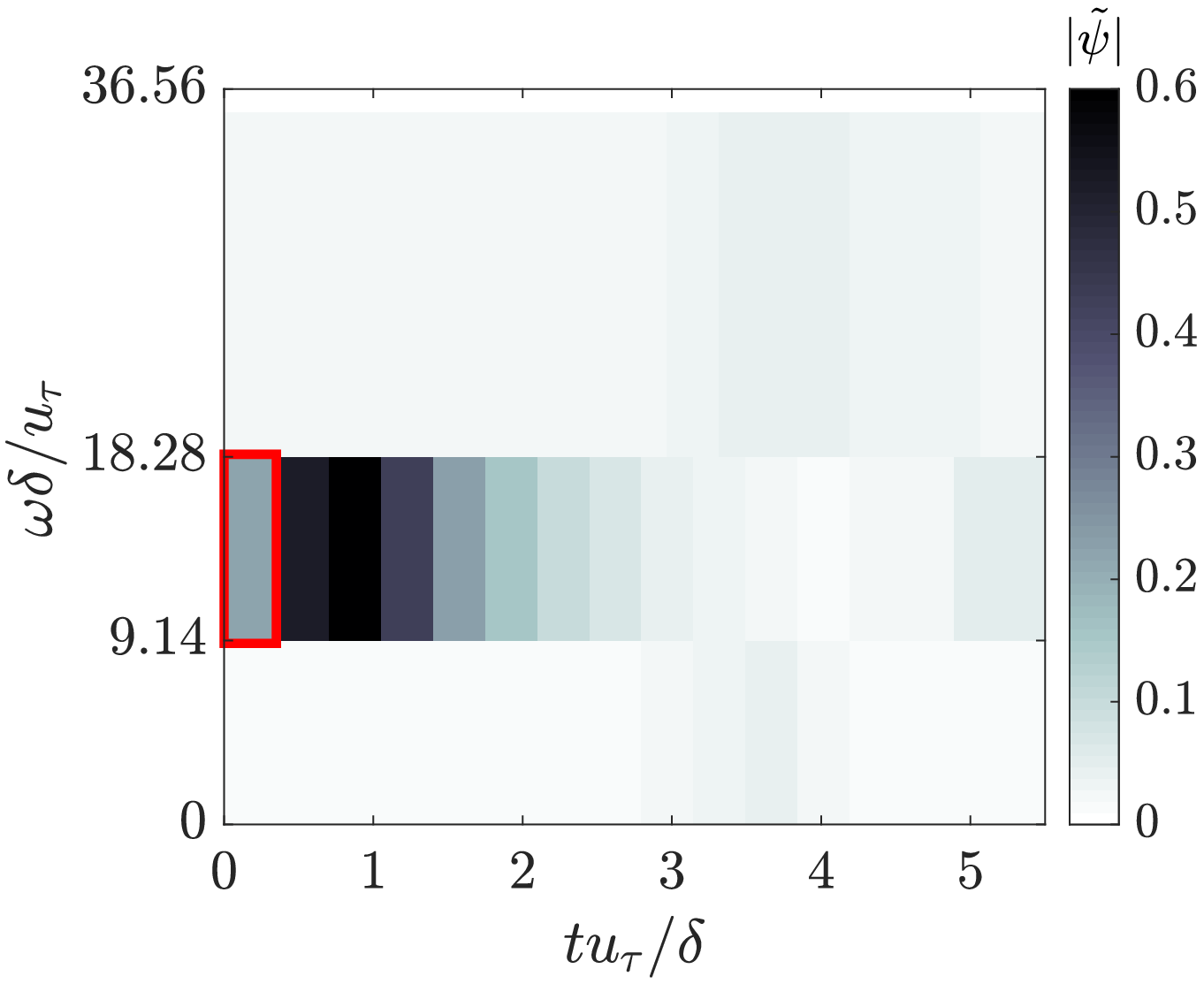}
    \caption{}
\end{subfigure}
\hspace{0.3cm}
\begin{subfigure}{0.445\textwidth}
\includegraphics[width=\linewidth]{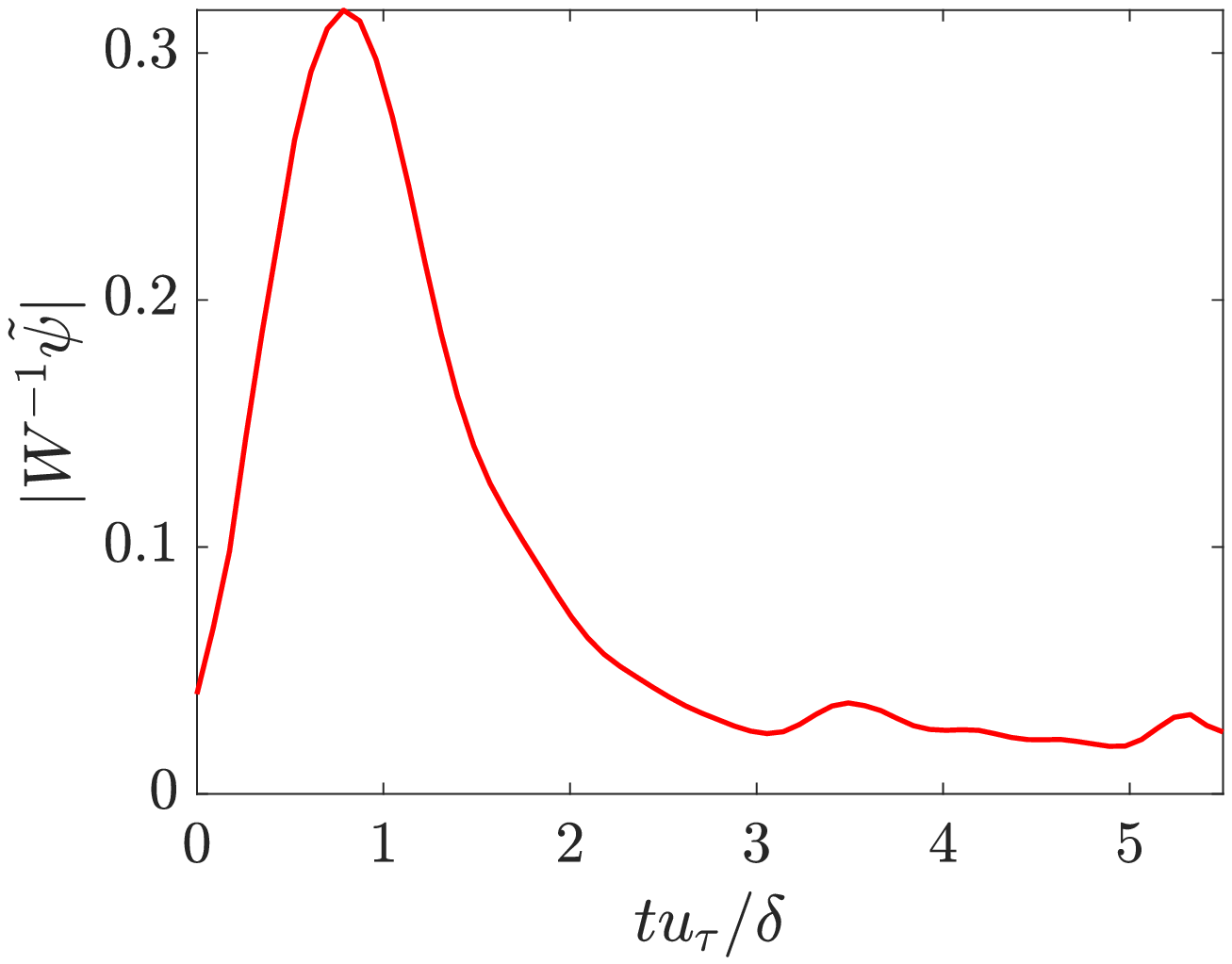}
\caption{}
\end{subfigure}
\caption{(a) Magnitude of the principal response mode of $\tilde{\mathcal H}$ for channel flow at $Re_\tau = 186$ under transient forcing, in the frequency-time plane. The forcing window is highlighted in red. (b) Magnitude of the inverse wavelet-transformed principal response mode as a function of time.}
\label{fig:results}
\end{figure}

The added advantage of the wavelet-based method lies in its ability to preserve temporal localization, which lets us formulate a time-scale sensitive resolvent for transient flows even when the mean profiles are statistically stationary. 
For this, we use the windowed wavelet-based resolvent analysis framework from \S\ref{sec:windowing}, but setting the windowing of the response modes, $\mathcal{C}$, to the identity matrix.
\begin{comment}
We can reformulate a resolvent map between inputs at a specific time shift and scale,
%
\begin{equation}
\left[\begin{array}{c}\tilde{u}_1(k_1,k_3)\\ \tilde{u}_2(k_1,k_3)\\ \tilde{u}_3(k_1,k_3)\\ \tilde{p}(k_1,k_3)\end{array}\right] = \tilde{\mathcal{H}}(k_1,k_3) \mathcal B \left[\begin{array}{c}\tilde{f}_1(k_1,k_3)\\ \tilde{f}_2(k_1,k_3)\\ \tilde{f}_3(k_1,k_3)\\ 0 \end{array}\right],
\label{eq:windowed}
\end{equation}
%
where $\mathcal B$ is a windowing matrix \cite{jeun2016, kojima2020} that selects for a subset of the full input states. 
\end{comment}
We then calculate the resolvent modes of $\tilde{\mathcal H}\mathcal B$. The principal forcing mode will lie in the span of a subset of the complete wavelet basis, as selected by $\mathcal B$. By picking a particular wavelet scale and shift via $\mathcal B$, we can restrict the forcing to a chosen frequency range, mostly localized in a particular time range.

To demonstrate this capability, we study the effect of a time-localized forcing on the buffer layer ($x_2^+ \in [15, 40]$). We choose $\mathcal{B}$ such that the forcing modes are limited to a single wavelet with the frequency range containing the critical layer corresponding to the buffer layer, i.e., $[U(x_2^+ = 15),U(x_2^+ = 40)] = [\omega_{\min}, \omega_{\max}]/k_1$ (Fig. \ref{fig:buffer}).
For the current case, we have $\omega_{\min} = 12.52 {u_\tau}/{\delta}$ and $\omega_{\max} = 17.17 {u_\tau}/{\delta}$. We pick a temporal resolution of $N_t=64$, and a time horizon of $T = 5.5 {\delta}/{u_\tau}$. We project the flow variables onto Shannon wavelets that cover frequency intervals $(2^{-i-1}{N_t}/{T})([-2\pi, -\pi] \cup [\pi, 2\pi])$, for $i=0, 1, \cdots 5$. The windowing matrix $\mathcal B$ is chosen to restrict the forcing to the unshifted wavelet ($\tau = 0$, i.e., centered at $t=0$) that has been scaled to cover $[9.14, 18.28]{u_\tau}/{\delta}$ (i.e. $i=1$).

\begin{figure}
\centering
\begin{subfigure}{0.8\textwidth}
    \includegraphics[width=\linewidth]{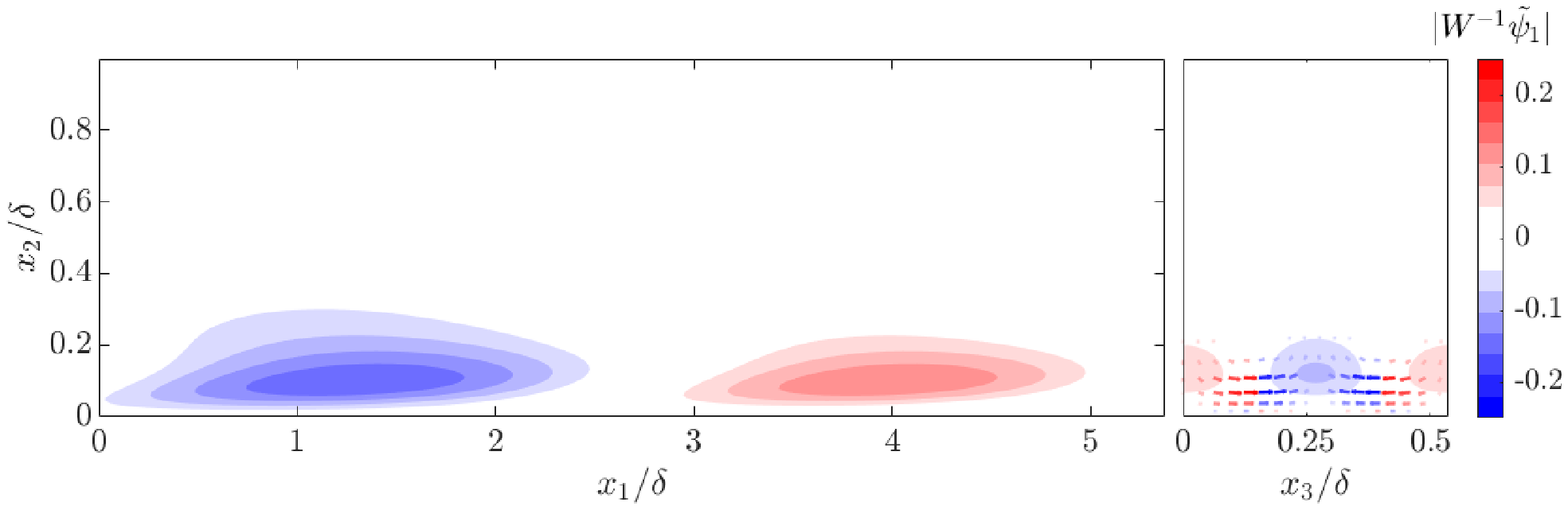} %17
    \caption{}
    %\vspace{-1cm}
    %19
    \includegraphics[width=\linewidth]{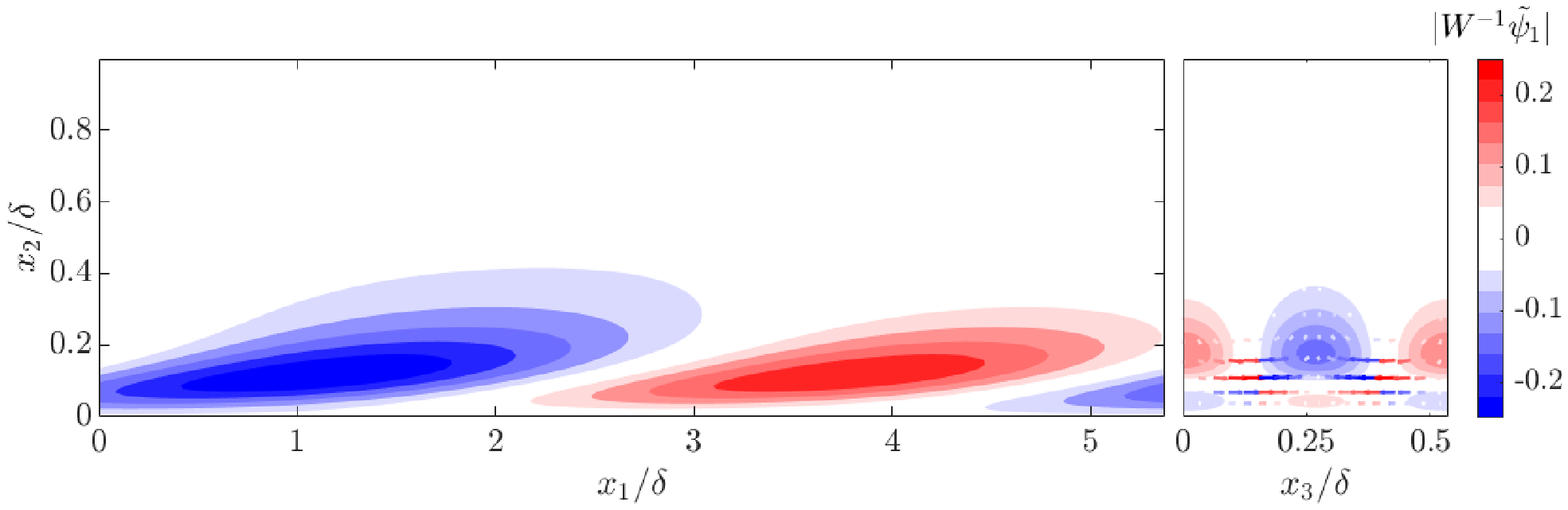}
    \caption{}
    %22
    \includegraphics[width=\linewidth]{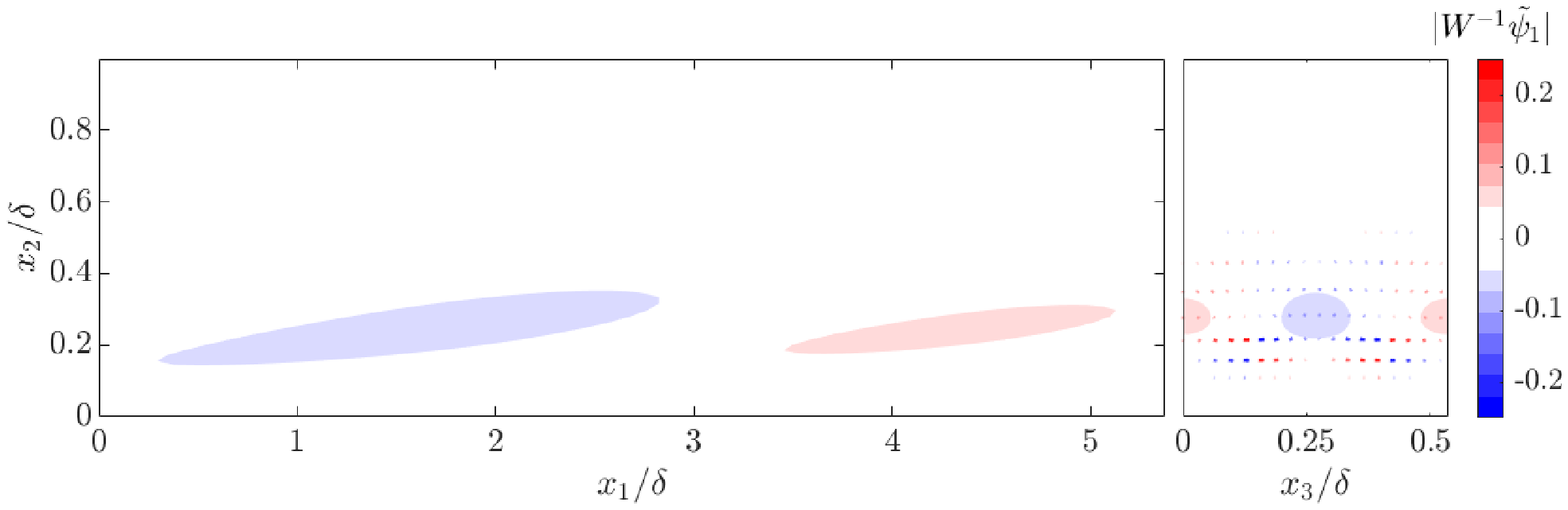}
    \caption{}
\end{subfigure}\qquad
%\begin{subfigure}[b]{0.01\textwidth}
%    \includegraphics[height=17cm]{arrow.jpg}
%\end{subfigure}

\caption{Principal response mode of the wavelet-based resolvent for the channel flow at $Re_\tau=186$ and under transient forcing, shown at (a) $t u_\tau/\delta = 0.26$, (b) $t u_\tau/\delta = 0.79$, and (c) $t u_\tau/\delta = 1.92$. The arrows in the right panels are colored according to the sign of $\sqrt{|\tilde{\psi_2}|^2 + |\tilde{\psi_3}|^2} sgn(\tilde{\psi_1})$ to show the intensity of the streamwise rolls relative to the streamwise streaks.}

\label{fig:streaks}

\end{figure}

The resulting principal response mode is confined to the frequency band determined by the forcing, as shown in Fig. \ref{fig:results}(a), which is expected since the time scales are decoupled for statistically stationary flows. The response mode additionally varies in time; the magnitude of the mode growth as a function of time is shown in Fig. \ref{fig:results}(b). The principal response mode peaks at a time $t u_\tau/\delta = 0.79$ before decaying. This transient growth can be explained by the non-normality of the linearized system \cite{schmid2000stability}. The response modes at three different time shifts are shown in Fig. \ref{fig:streaks}. The modes are predominantly in the streamwise direction, forming alternating low- and high-speed streamwise streaks. Although not shown, the principal forcing mode only has a small contribution from the streamwise direction and is in the form of streamwise rolls, in line with the linear theory \cite{orr1907,landahl1975, howlinear} and the self-sustaining process of wall turbulence \cite{jimenez1991,hamilton1995,jimenez1999,waleffe1997,schoppa2002,farrell2017,bae2020resolvent}. %While the turbulent channel is a statistically stationary flow, we note that the transient behavior of the resolvent modes cannot be captured by Fourier-based resolvent analysis.

\section{Application to non-stationary flow} \label{sec:app_time}

We now apply wavelet-based resolvent analysis to problems with a time-varying mean flow. In particular, we study the turbulent Stokes boundary layer and a turbulent channel flow with a sudden lateral pressure gradient. The Stokes boundary layer is a purely oscillatory flow in time, and thus, Fourier-based resolvent analysis still may be used \cite{Padovan2020}. However, in the case of the temporally-changing channel flow, the flow is truly unsteady, and a Fourier transform in time will not be applicable. 

\subsection{Turbulent Stokes boundary layer}

\begin{figure}
    \centering
    \begin{subfigure}{0.45\textwidth}
    \includegraphics[width=\linewidth]{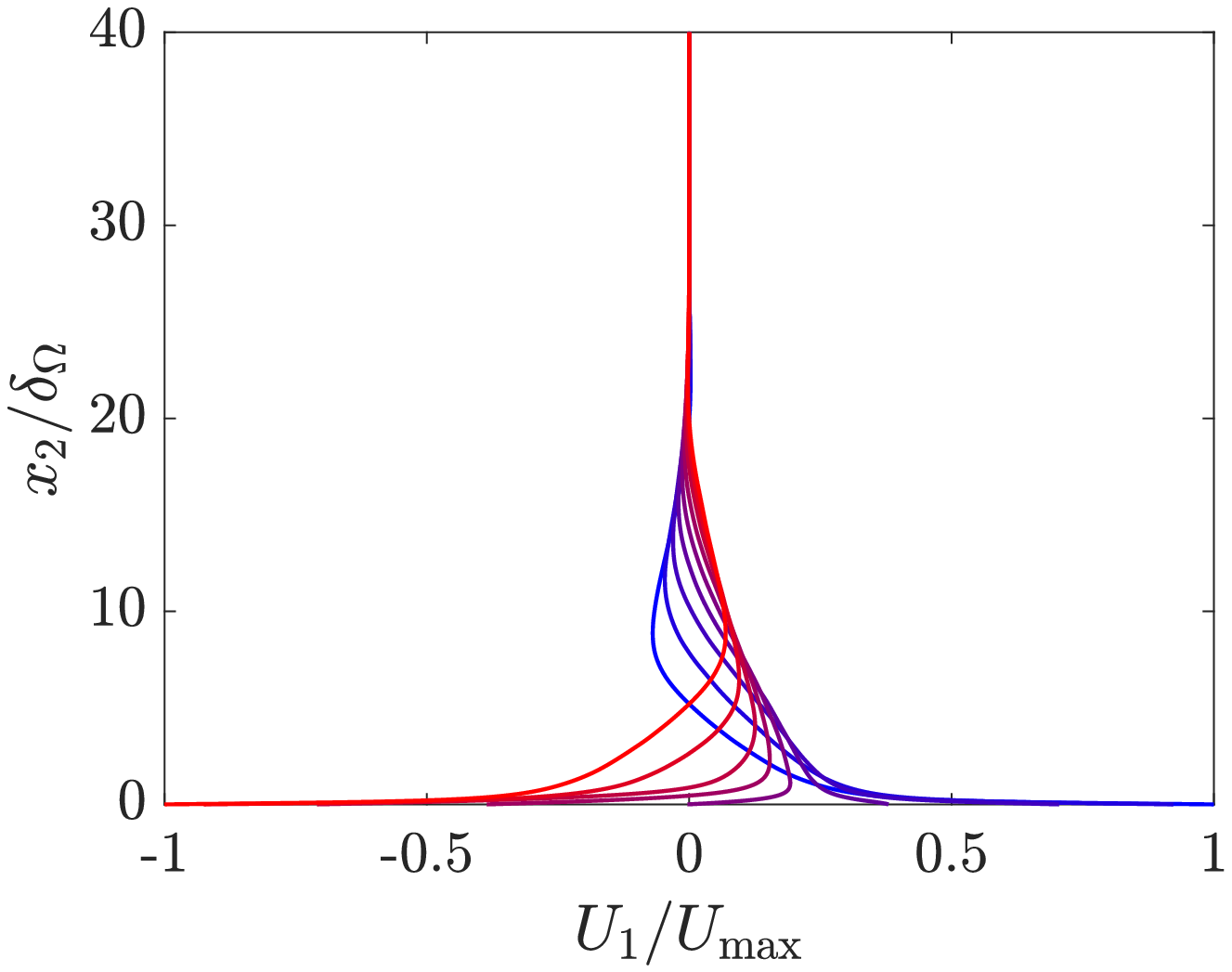}
    \caption{}
    \end{subfigure}
    \hspace{0.2cm}
    \begin{subfigure}{0.46\textwidth}
    \includegraphics[width=\linewidth]{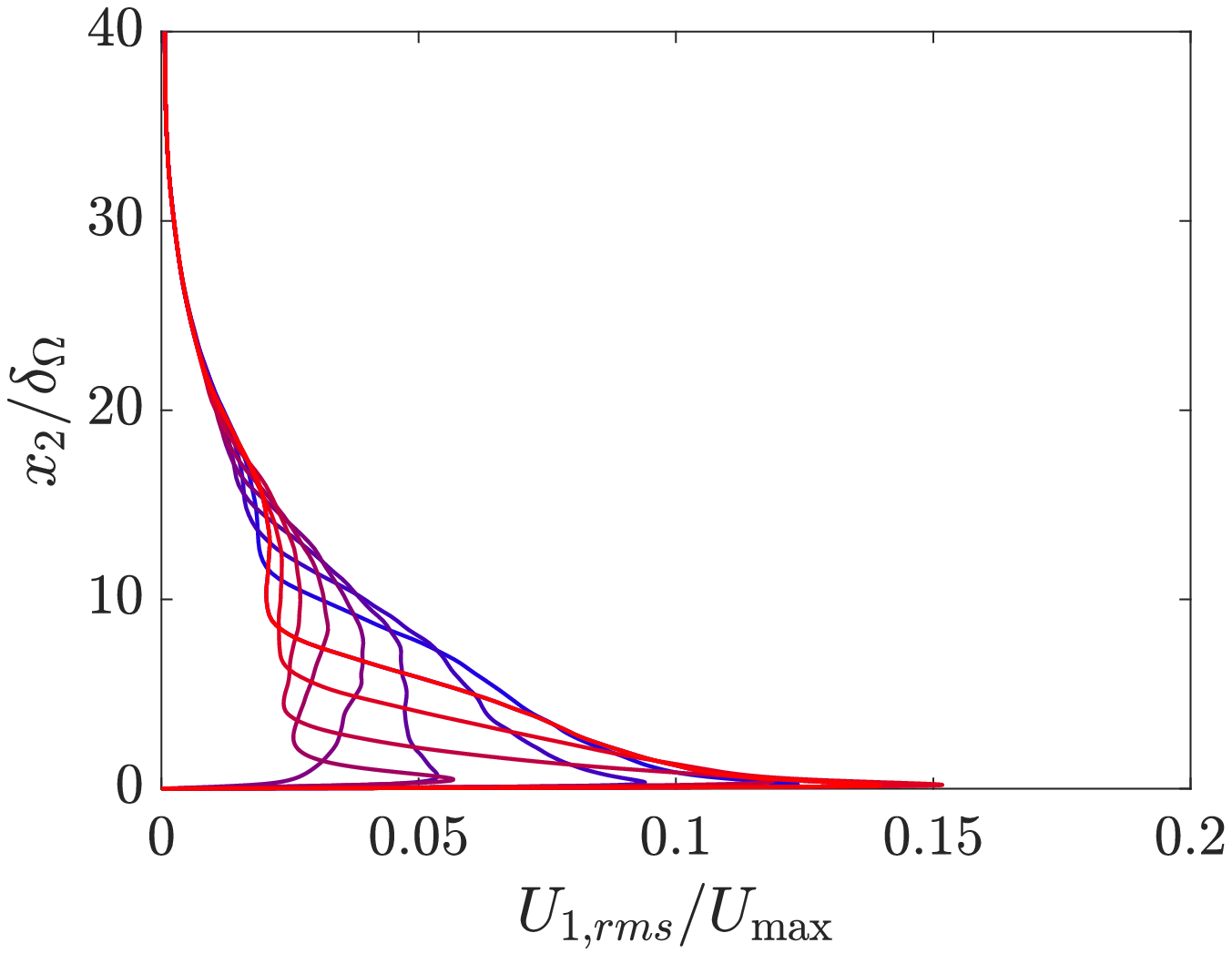}
    \caption{}
    \end{subfigure}
    \caption{(a) Mean streamwise velocity profile and (b) streamwise r.m.s. velocity at from $t\Omega = 0$ (blue) to $t\Omega = \pi$ (red). The profiles shown are at $t\Omega =0, \, \pi/8,\,\pi/4, \, 3\pi/8, \, \pi/2, \, 5\pi/8, \, 3\pi/4 ,\, 7\pi/8, \pi$.}
    \label{fig:StokesMean}
\end{figure}

The Stokes boundary layer is simulated through a channel flow with the lower and upper walls oscillating in tandem at a velocity of $U_w(t) = U_{\max}$ cos$(\Omega t)$ with no imposed pressure gradient. The relevant nondimensional number is the Reynolds number $Re_\Omega = U_{\max} \delta_\Omega / \nu$, where $\delta_\Omega = \sqrt{2\nu/\Omega}$ denotes the Stokes boundary layer thickness. For the current case, we consider $Re_\Omega = 1500$, which lies within the intermittently turbulent regime \cite{hino1976,akhavan1991,verzicco1996,vittori1998,costamagna2003}. This problem has been well-studied numerically and experimentally in the literature \cite{hino1976,spalart1989direct,jensen1989turbulent,akhavan1991,verzicco1996,vittori1998,costamagna2003,kerczekDavis,sarpkaya1993coherent,blondeauxVittori,carstensen,ozdemir}. The same numerical solver used to simulate the turbulent channel flow is used to generate the statistics, with modifications in the boundary condition to accommodate the oscillating walls. The domain size of the channel for the DNS is given by $6\pi \delta_\Omega \times 80 \delta_\Omega \times 3\pi \delta_\Omega$ and discretized using $64$, $385$ and $64$ points in each direction. We compute the mean velocity profiles by averaging in homogeneous directions and phase. Fig.~\ref{fig:StokesMean} shows the mean and the root-mean-square (rms). velocity profiles at three different temporal locations. Note that $U_1(t\Omega + \pi) = -U(t\Omega)$ and $U_{i,rms}(t\Omega + \pi) = U_{i,rms}(t\Omega)$. We observe that the turbulent energy peaks near the wall at $t\Omega = 2.65$ and propagates away from the wall in time.

%\begin{figure}{}

%\begin{subfigure}[b]{0.335\textwidth}
%    \includegraphics[width=\linewidth]{meanProf1.eps} 
%    \caption{}
%\end{subfigure}
%\begin{subfigure}[b]{0.3\textwidth}
%    \includegraphics[width=\linewidth]{meanProf2.eps} 
%    \caption{}
%\end{subfigure}
%\begin{subfigure}[b]{0.3\textwidth}
%    \includegraphics[width=\linewidth]{meanProf3.eps} 
%    \caption{}
%\end{subfigure}

%\begin{subfigure}[b]{0.34\textwidth}
%    \includegraphics[width=\linewidth]{rms1.eps} 
%    \caption{}
%\end{subfigure}
%\begin{subfigure}[b]{0.3\textwidth}
%    \includegraphics[width=\linewidth]{rms2.eps} 
%    \caption{}
%\end{subfigure}
%\begin{subfigure}[b]{0.3\textwidth}
%    \includegraphics[width=\linewidth]{rms3.eps} 
%    \caption{}
%\end{subfigure}
%\caption{Mean streamwise velocity profile in the lower half of the channel at (a) $t\Omega = 0$, (b) $t\Omega = \pi/2$, and  (c) $t\Omega = \pi$. The r.m.s. streamwise (black), wall-normal (red), and spanwise (blue) velocity profiles at (d) $t\Omega = 0$, (e) $t\Omega = \pi/2$, and  (f) $t\Omega = \pi$.}

%\label{fig:StokesMean}

%\end{figure}

To construct the resolvent operator, we first choose the spatial scales for the homogeneous directions. Using the DNS data, we calculate the streamwise energy spectrum at $x_2/\delta_\Omega = 1.43$ and $t\Omega = 2.51$, which correspond to the peak spatio-temporal location of the $U_{1,rms}$. The most energetic streamwise and spanwise scales at that location are given by $k_1\delta_\Omega = 0.67$ and $k_3\delta_\Omega = 2.67$, which we choose as the streamwise and spanwise scales for the resolvent operator. We use $N_{x_2} = 140$ and $N_t = 780$ to solve the discrete system. We construct $D_t$ and $D_{x_2}$ as sparse finite difference matrices. We additionally choose the time-derivative operator $D_t$ to be circulant to enforce periodicity in time. We compute the modes for the half-channel and modify $D_{x_2}$ to enforce a no-slip and no-penetration boundary condition at the wall, and a free-slip and no-penetration boundary condition at the centerline.

\begin{figure}
    \centering
    \begin{subfigure}{0.59\textwidth}
    \includegraphics[width=\textwidth]{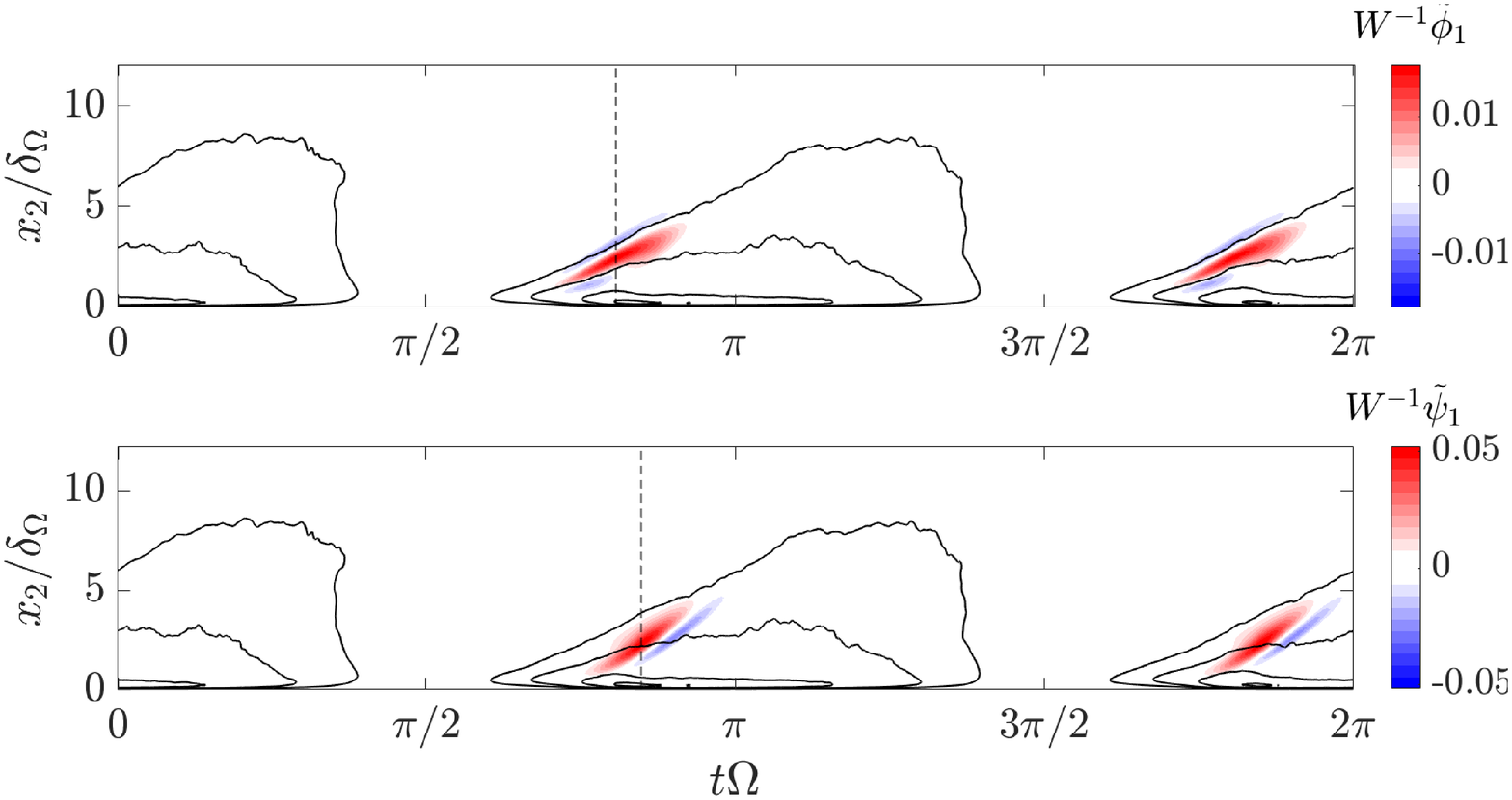}
    \caption{}
    \end{subfigure}
    \begin{subfigure}{0.39\textwidth}
    \includegraphics[width=\textwidth]{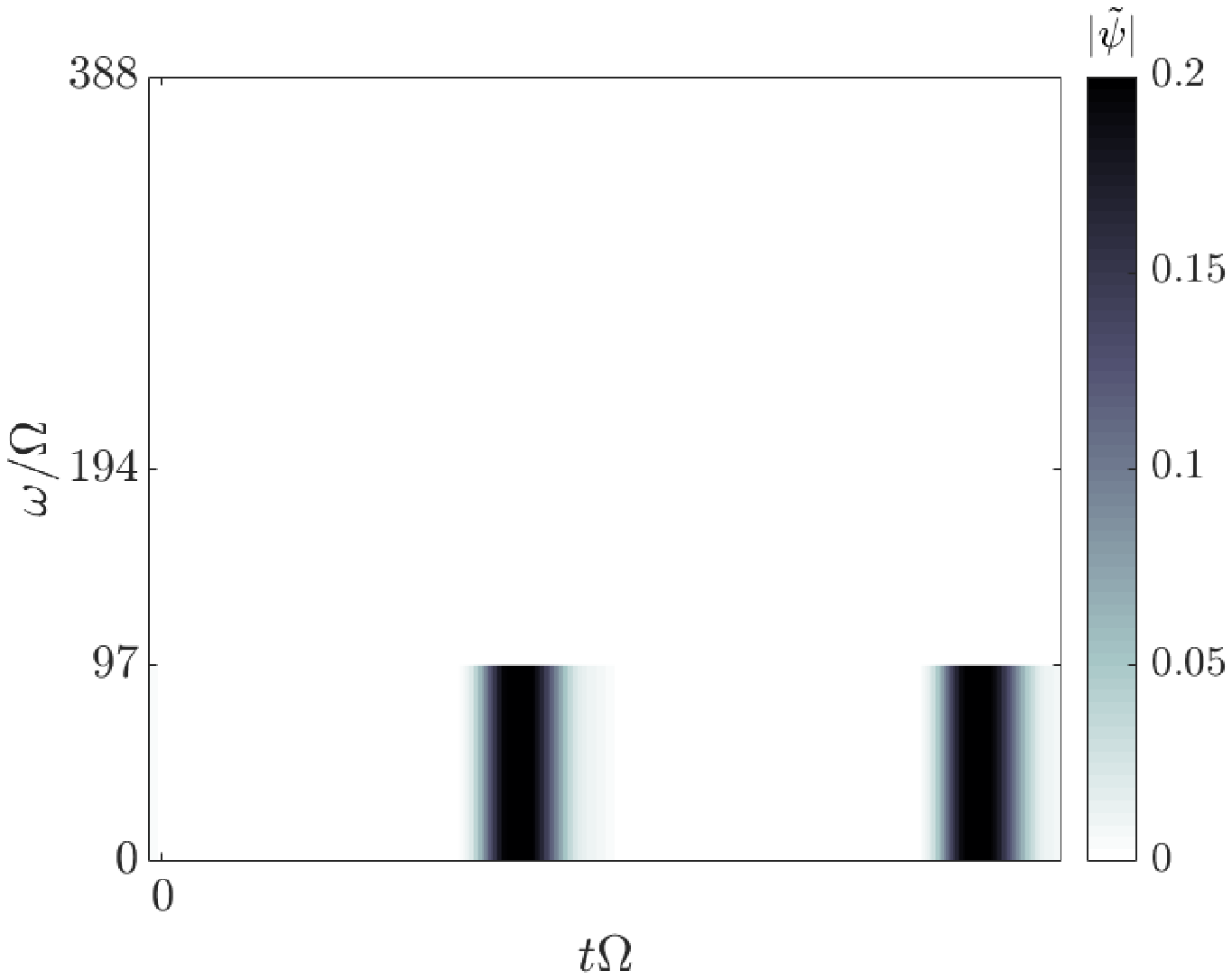}
    \caption{}
    \end{subfigure}
    \caption{(a) Streamwise velocity component of the principal resolvent input (top) and output (bottom) modes of the Stokes boundary layer. The black contour lines are $U_{1, rms}$ with the levels indicating $30\%, 50\%, 75\%, 95\%$ of its maximum value. The vertical dashed lines show the location of the amplitude peak for the input mode ($t\Omega = 2.54$) and output mode ($t\Omega = 2.67$). (b) Magnitude of the principal response mode of $\tilde{H}$ for the Stokes boundary layer in the frequency-time plane.}
    \label{fig:stokesModes}
\end{figure}

\begin{figure}{}

    \centering
    \begin{subfigure}[b]{0.45\textwidth}
    \includegraphics[width=\linewidth]{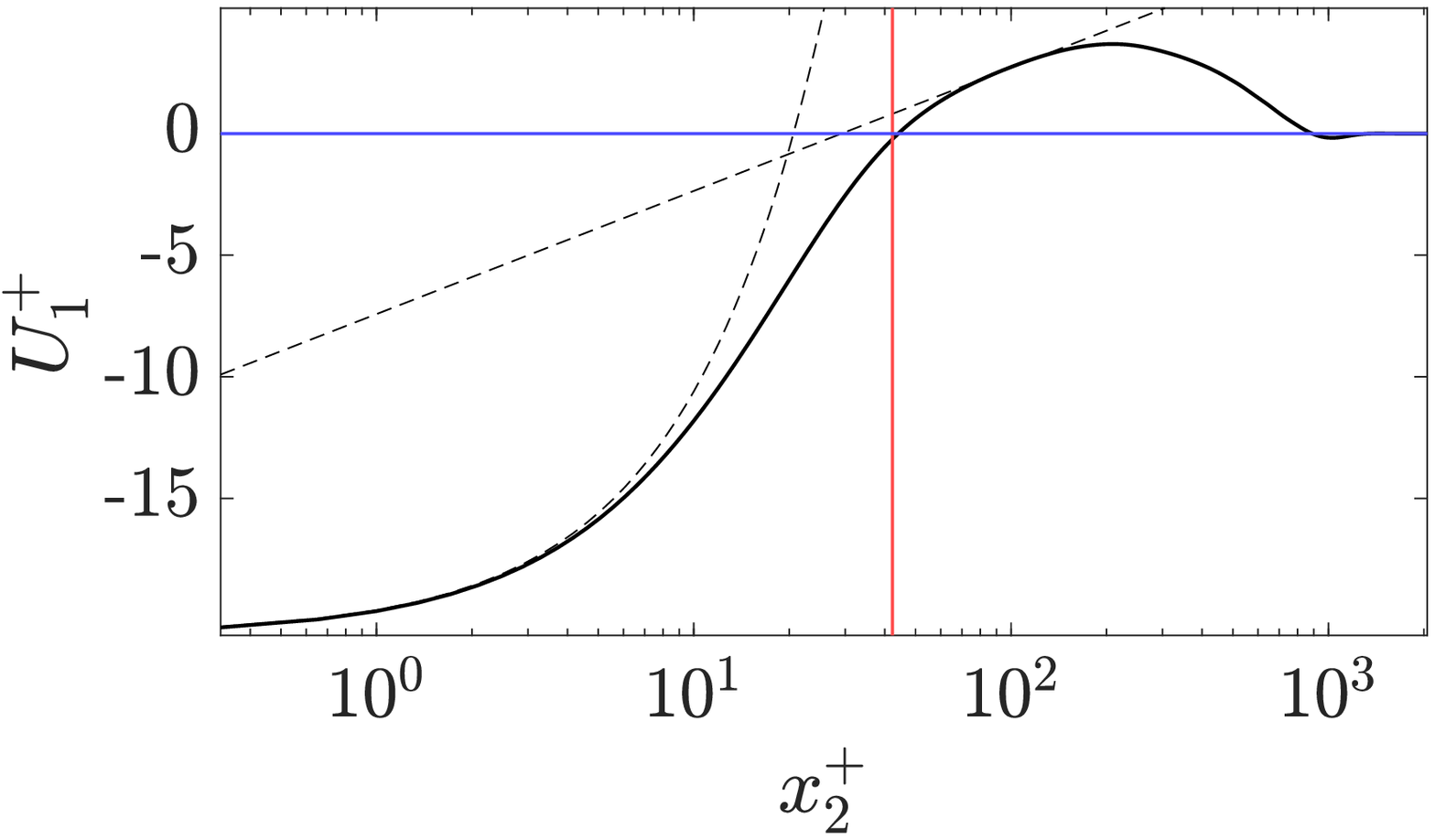}
    \caption{}
    \end{subfigure}  
    \hspace{0.2cm}
    \begin{subfigure}[b]{0.45\textwidth}
    \includegraphics[width=\linewidth]{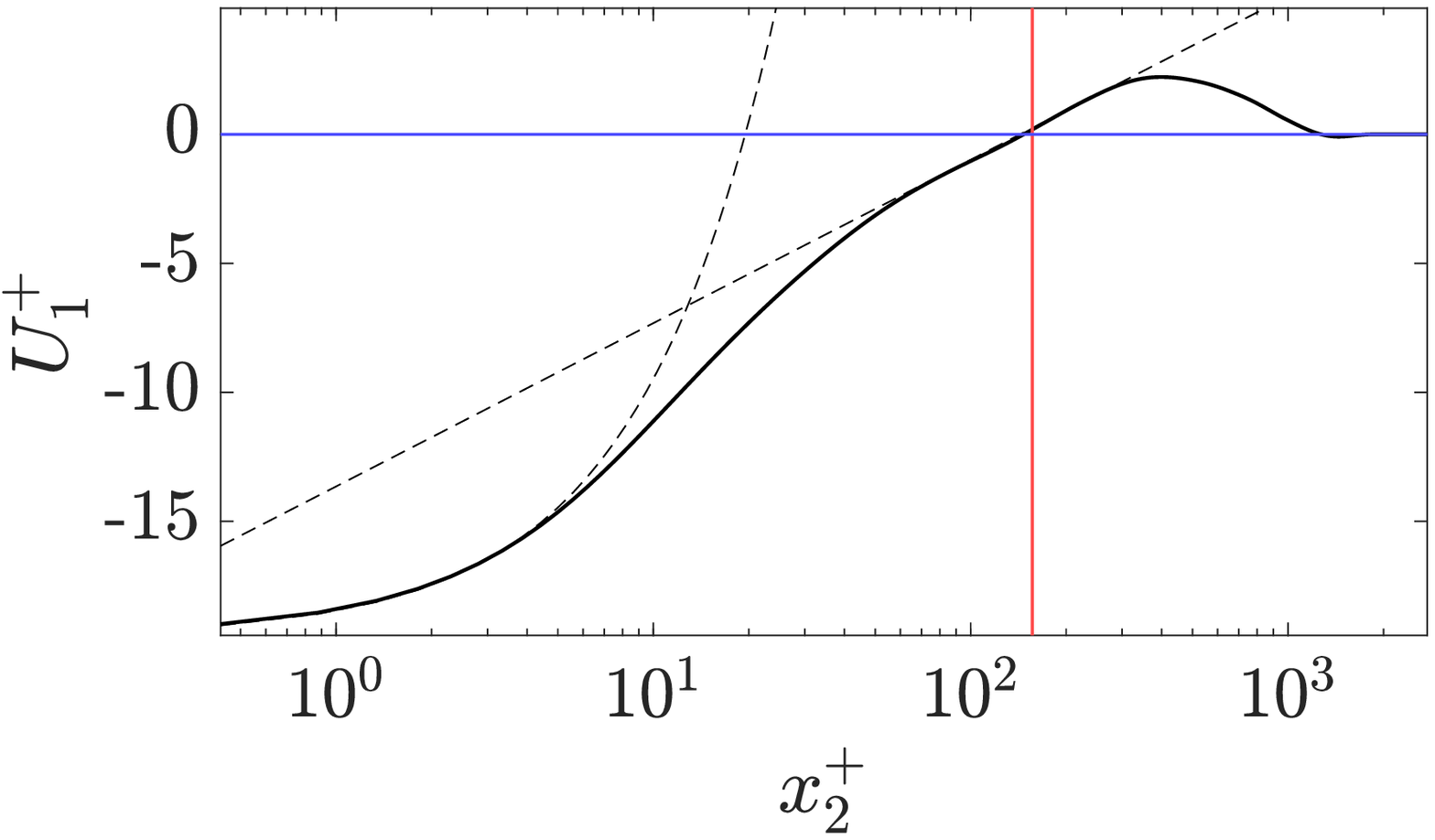}
    \caption{}
    \end{subfigure}
    
    \caption{Mean streamwise velocity profiles normalized by $u_\tau(t\Omega)$ at (a) $t\Omega=2.35$ and (b) $t\Omega=2.67$, which correspond to 10\% and 100\% of the maximum intensity of the principal response mode. The vertical red lines show the wall-normal location of the instantaneous peak of the streamwise component of the output mode, the dashed black lines show the viscous and logarithmic layers, and the horizontal blue line marks where the velocity is zero.}
    \label{fig:StokesLogLayer}
\end{figure}

The principal input and output modes corresponding to the chosen spatial scales and boundary conditions in time are shown in Fig. \ref{fig:stokesModes}(a). We observe that the principal input and output modes are synchronized with the peaks in $U_{1, rms}$. The energy propagation towards the center occurs at a similar rate for the resolvent modes as for the DNS results. This suggests that the energy amplification in the Stokes boundary layer can partially be explained by the optimal linear mechanism, as for turbulent channel flow \cite{howlinear}. Moreover, we observe that the principal input modes precede the principal output modes with time delay, and thus a transient growth mechanism can be explained through the wavelet-based resolvent modes. This is in line with a physical interpretation of the modes in which the input modes `cause' the output modes. In Fig. \ref{fig:stokesModes}(b), we show the magnitude of the response modes in the frequency-time plane, which highlights that the bulk of the response is limited to a single wavelet scale and a few time shifts. Using the windowed resolvent operator to restrict our forcing and our response to the first wavelet scale, we indeed obtain nearly identical modes as in \ref{fig:stokesModes} while reducing the effective dimension of our resolvent operator by a factor of four.

Fig. \ref{fig:StokesLogLayer} shows that the peak amplification occurs at the location of zero mean streamwise velocity. This is in line with the fact that Fourier-based resolvent modes are often centered around a critical layer \cite{McKeon2010}, where the critical layer $x_2 = x_2^c$ is defined as $U_1(x_2^c) = c$ and $k_x c = \omega$. For the Stokes boundary layer with the chosen length scales, the critical layer seems to occur at $U_1(x_2^c) = 0$, which corresponds to $\omega = 0$. For this time-periodic problem, we note that wavelet resolvent analysis should be equivalent to a harmonic resolvent analysis \cite{Padovan2020} that includes the interactions between all the resolved time scales. The wavelet resolvent modes should indeed map to the harmonic resolvent modes via an inverse wavelet transform and a Fourier transform in time. We expect that harmonic resolvent analysis will also reveal a peak amplification for waves at $\omega = 0$; we will conduct a harmonic resolvent analysis on this system and compare it with the wavelet-based resolvent modes in future works.

\subsection{Channel flow with sudden lateral pressure gradient}

\begin{figure}
    \centering
    \begin{subfigure}{0.45\textwidth}
        \includegraphics[width=\linewidth]{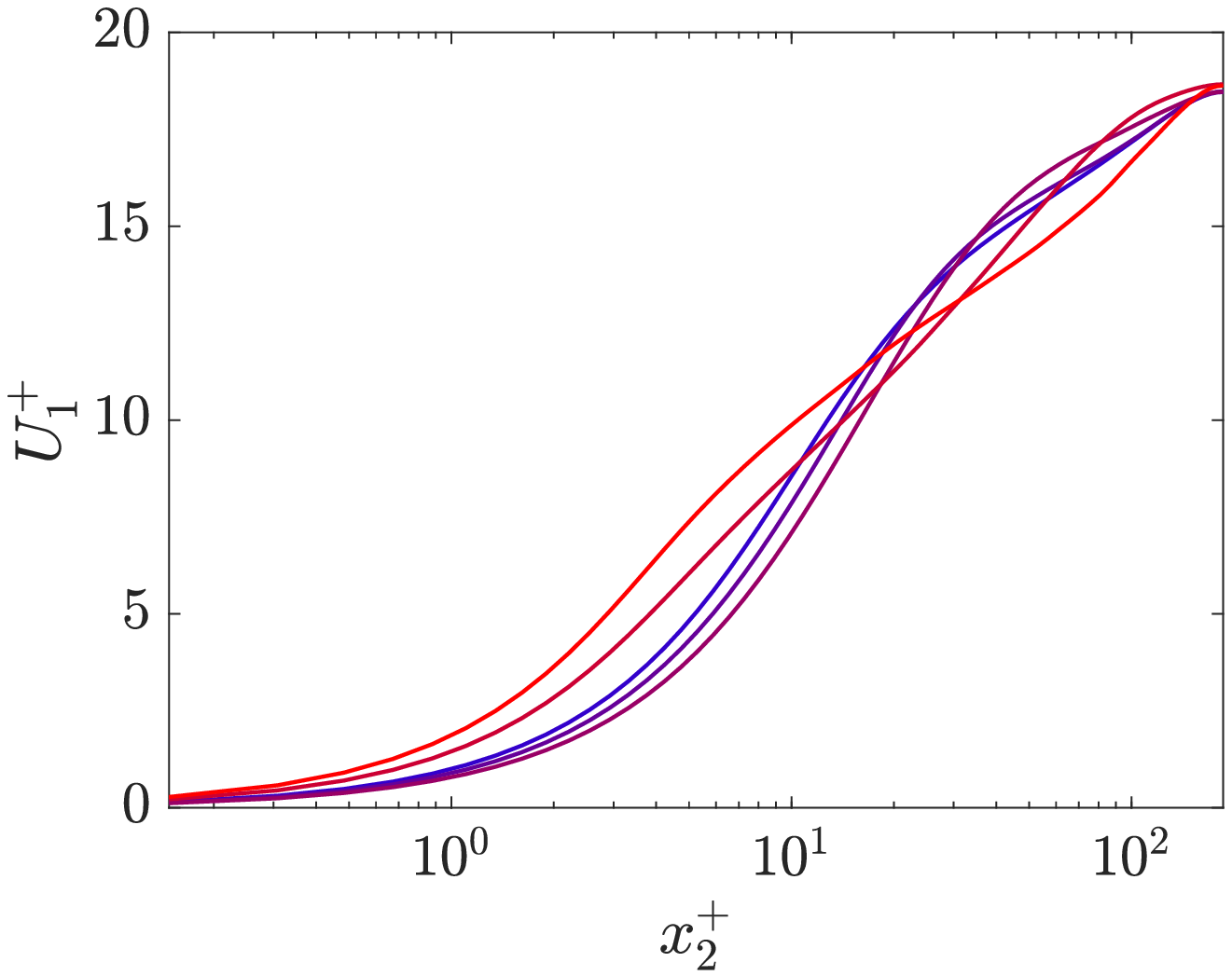}
        \caption{}
    \end{subfigure}
    \hspace{0.2cm}
    \begin{subfigure}{0.45\textwidth}
        \includegraphics[width=\linewidth]{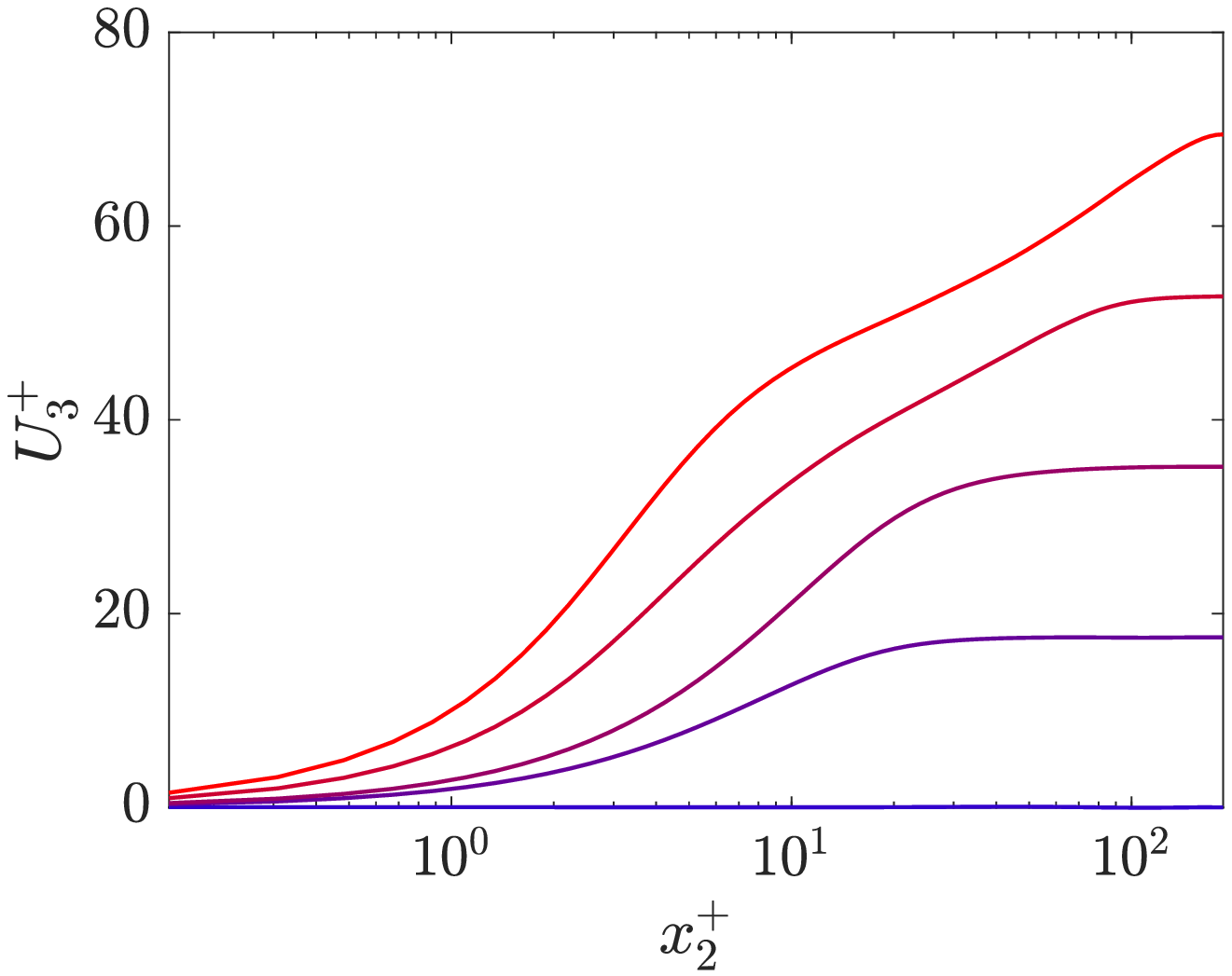}
        \caption{}
    \end{subfigure}
    \caption{Mean (a) streamwise and (b) spanwise velocity profile from $tu_{\tau, 0}/\delta = 0$ (blue) to $t u_{\tau, 0}/\delta = 2.34$ (red). The times shown are $tu_{\tau, 0}/\delta = 0, \, 0.58, \, 1.17,\, 1.76,\, 2.34$. Data taken from \cite{noneq}.}
    \label{fig:3DChannelProf}
\end{figure}

\begin{figure}
    \centering
    \begin{subfigure}{0.45\textwidth}
        \includegraphics[width=\linewidth]{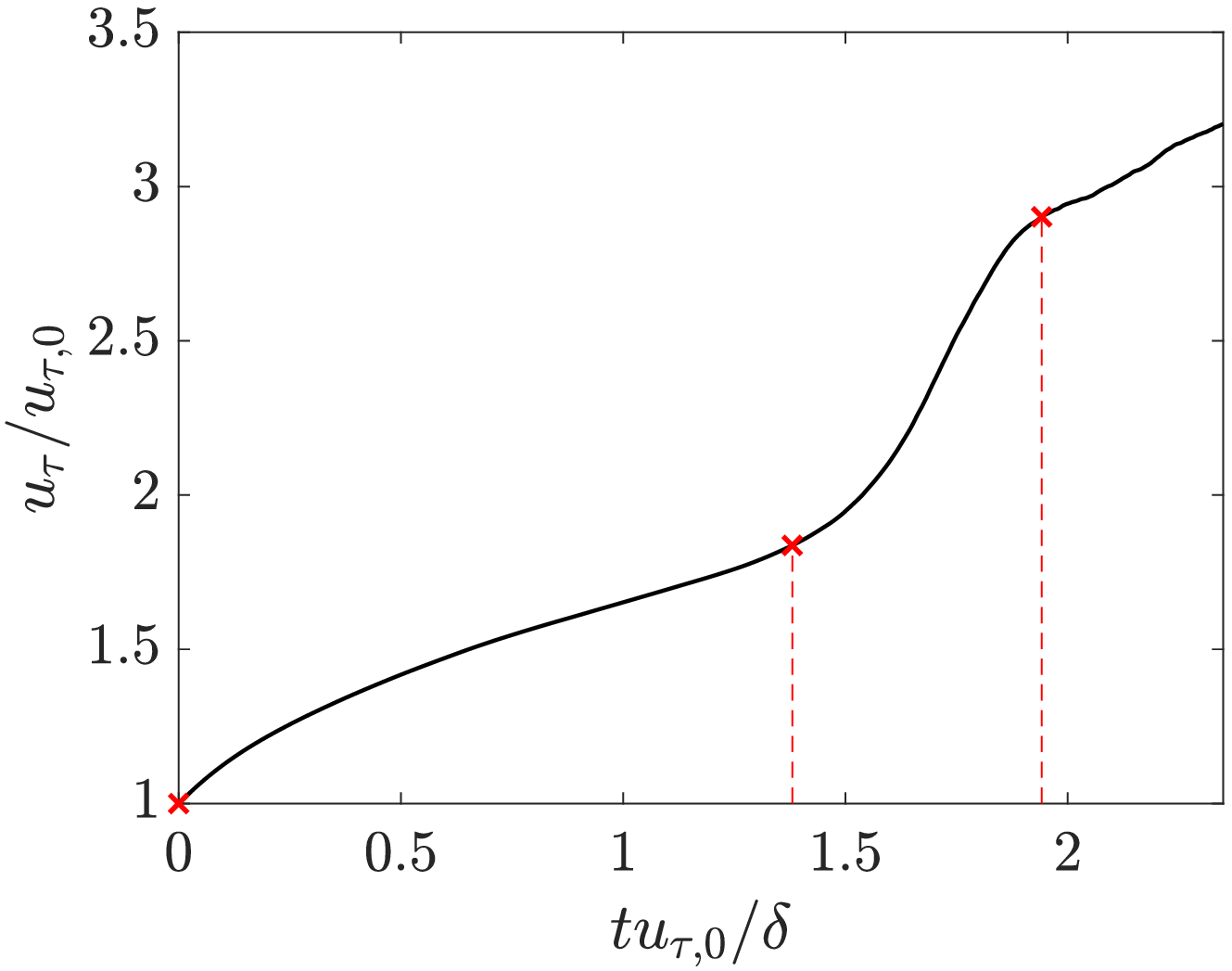}
        \caption{}
    \end{subfigure}
    \hspace{0.2cm}
    \begin{subfigure}{0.45\textwidth}
        \includegraphics[width=\linewidth]{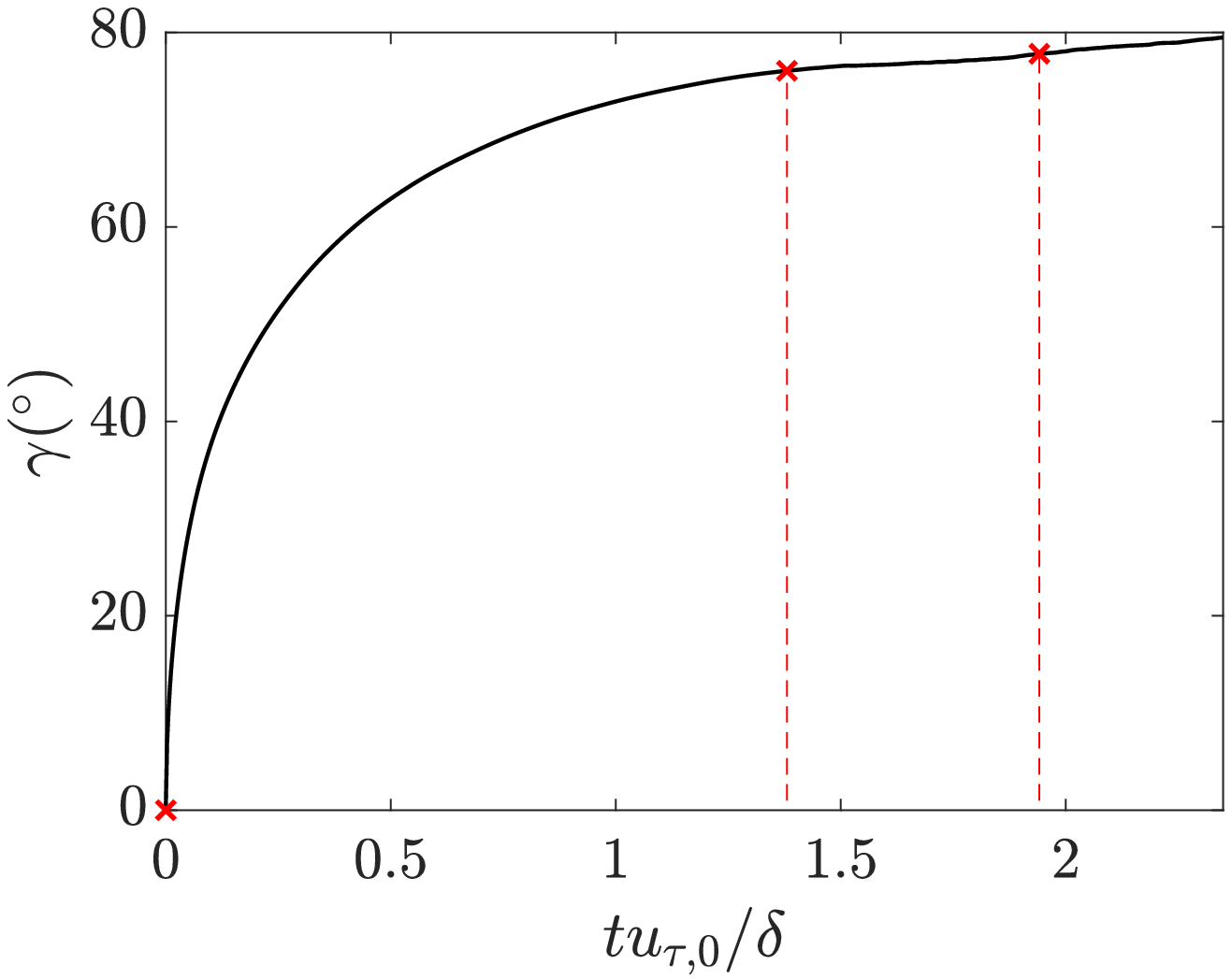}
        \caption{}
    \end{subfigure}
    \caption{(a) Friction velocity $u_{\tau}$ and (b) wall-shear stress angle $\gamma = \tan^{-1}(\tau_3 / \tau_1)$ as a function of time. Data taken from \cite{noneq}. The vertical dashed lines are at $t\Omega = 0, 1.38, 1.94$, and correspond to the choice of $\lambda_1^+$ and $\lambda_3^+$ for the modes plotted in Fig.\ref{fig:3DChannel}}
    \label{fig:utau_angle}
\end{figure}

Finally, we study a fully-developed turbulent channel flow at $Re_\tau = 186$ that is subjected to a sudden lateral pressure gradient $dP/dx_3 = \Pi dP/dx_1$ at $t=0$ with $\Pi = 30$ \cite{Moin1990, noneq}. This flow, commonly referred to as a three-dimensional (3D) channel flow, has an initial transient period dominated by 3D non-equilibrium effects. Eventually, the flow will reach a new statistically steady state with the mean flow in the $(dP/dx_1, dP/dx_3)$ direction parallel to the wall. In the transient period, the tangential Reynolds stress initially decreases before increasing linearly, with depletion and increase rate that scales as $\Pi x_2/\delta$ \citep{noneq}. 

The mean flow profiles are obtained from \cite{noneq} and have nonzero streamwise and spanwise components $U_1$ and $U_3$ (Fig. \ref{fig:3DChannelProf}) as well as nonzero wall-normal gradients of streamwise and spanwise components $dU_{1, 2}$ and $dU_{3, 2}$. The time domain of the simulation is $Tu_{\tau,0}/\delta = 2.34$, where $u_{\tau,0}$ is the initial friction velocity. To construct the discrete resolvent operator, we use a spatial resolution of $N_{x_2} = 65$ for the half-channel, and a finite difference matrix $D_{x_2}$ enforcing a no-slip and no-penetration boundary condition at the wall and a free-slip and no-penetration condition at the centerline. The boundary condition for the temporal finite difference operator $D_t$ is chosen to enforce a Neumann-type condition, $\partial_t(\cdot)|_{t=0} = \partial_t(\cdot)|_{t=T} = 0$. To reduce the impact of the boundary condition on the modes at $t=0$ we extend $U_1$ and $U_3$ to the time interval $tu_{\tau,0}/\delta \in [-0.58, 2.34]$ and assume $U_i(t \leq 0, y) = U_i(t = 0, y)$ and $dP/dx_3(t < 0) = 0$. When the modes are plotted, we only show the original time domain $tu_{\tau,0}/\delta \in [0, 2.34]$ and exclude the contribution from negative times. We use a temporal resolution of $N_t = 1000$ for the extended time frame. 

\begin{figure}
\centering
\begin{subfigure}{0.45\textwidth}
    \includegraphics[width=\linewidth]{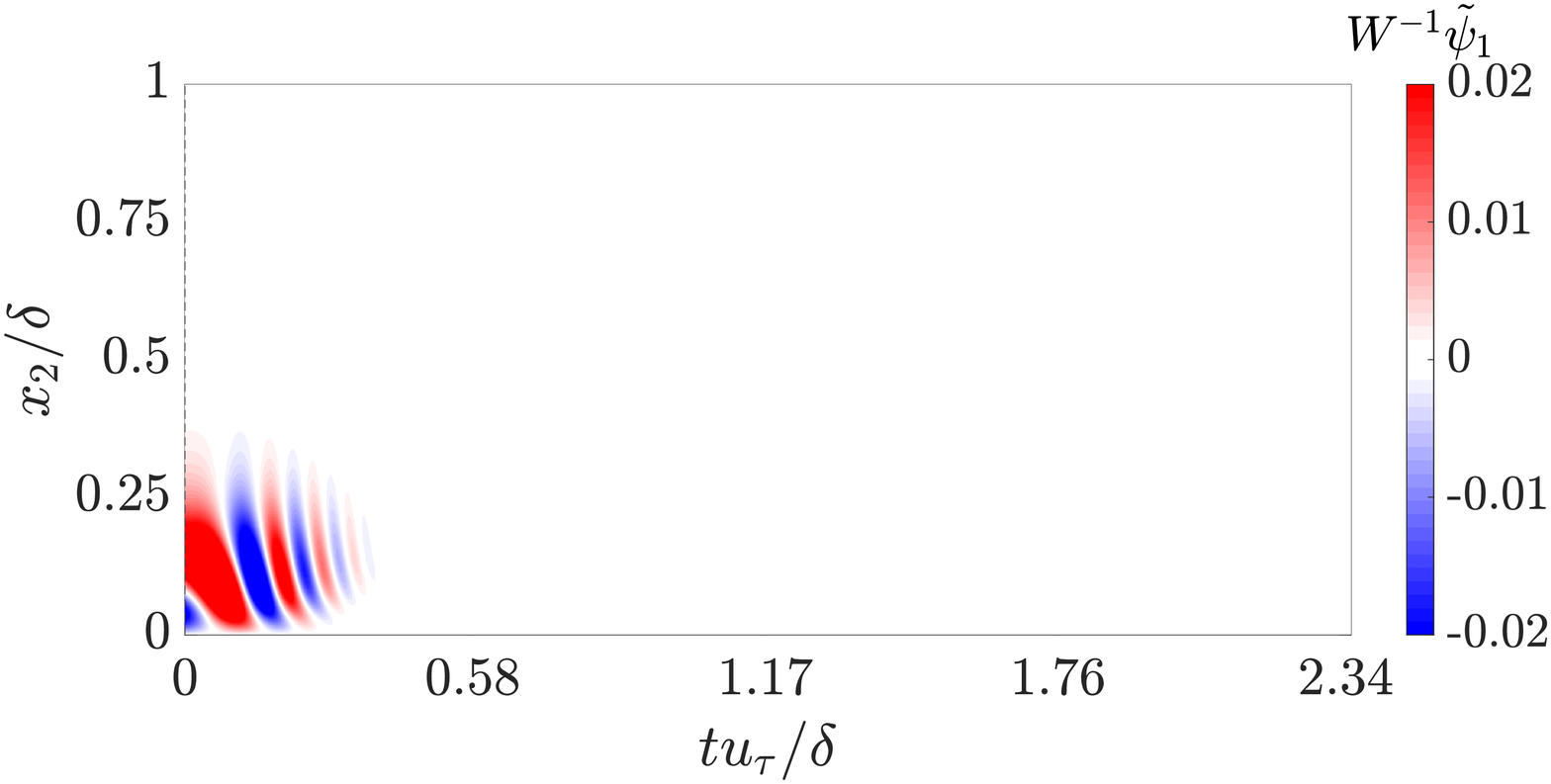}
    \caption{}
\end{subfigure}
\begin{subfigure}{0.45\textwidth}
    \includegraphics[width=\linewidth]{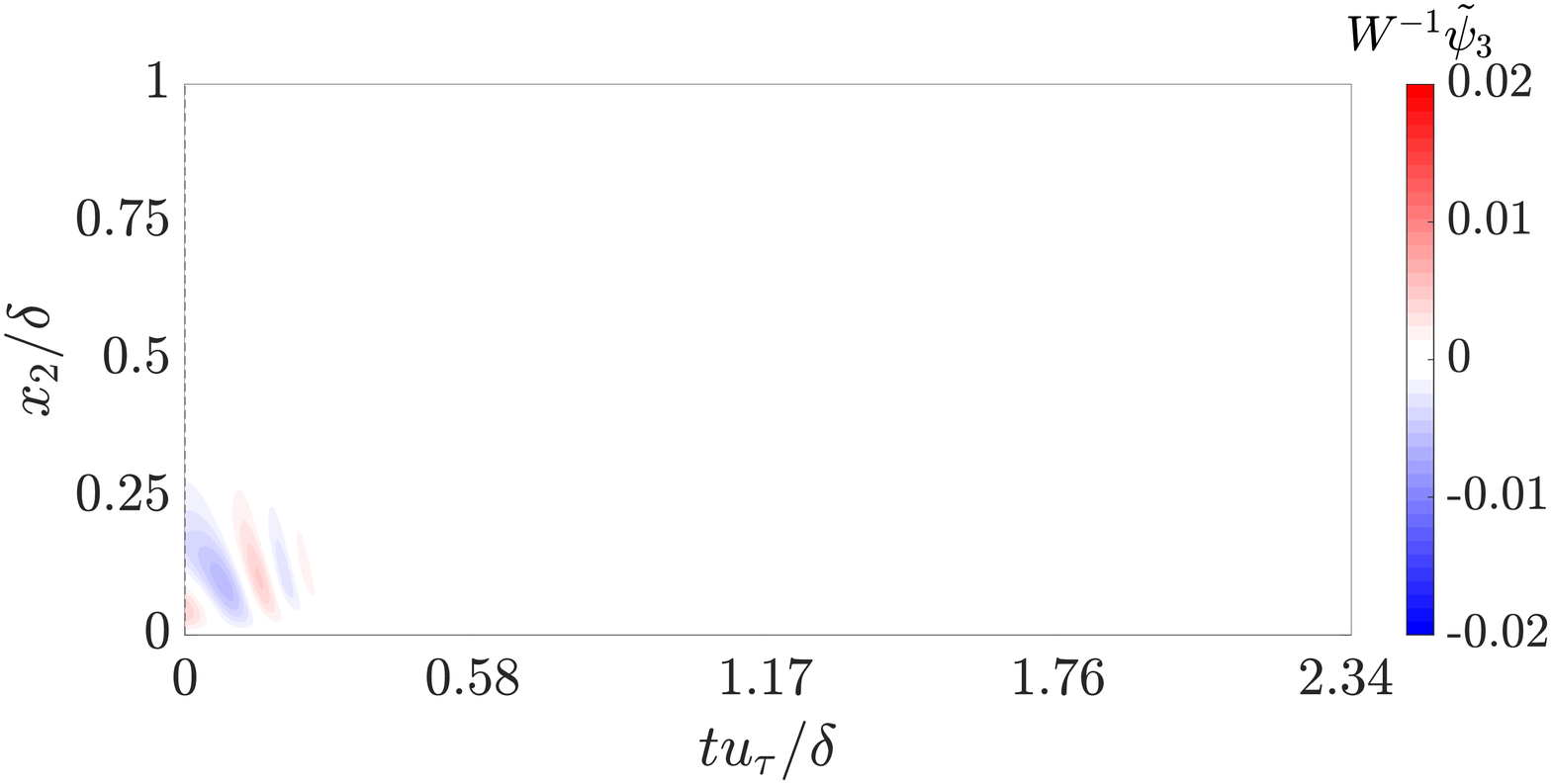}
    \caption{}
\end{subfigure}
\begin{subfigure}{0.45\textwidth}
    \includegraphics[width=\linewidth]{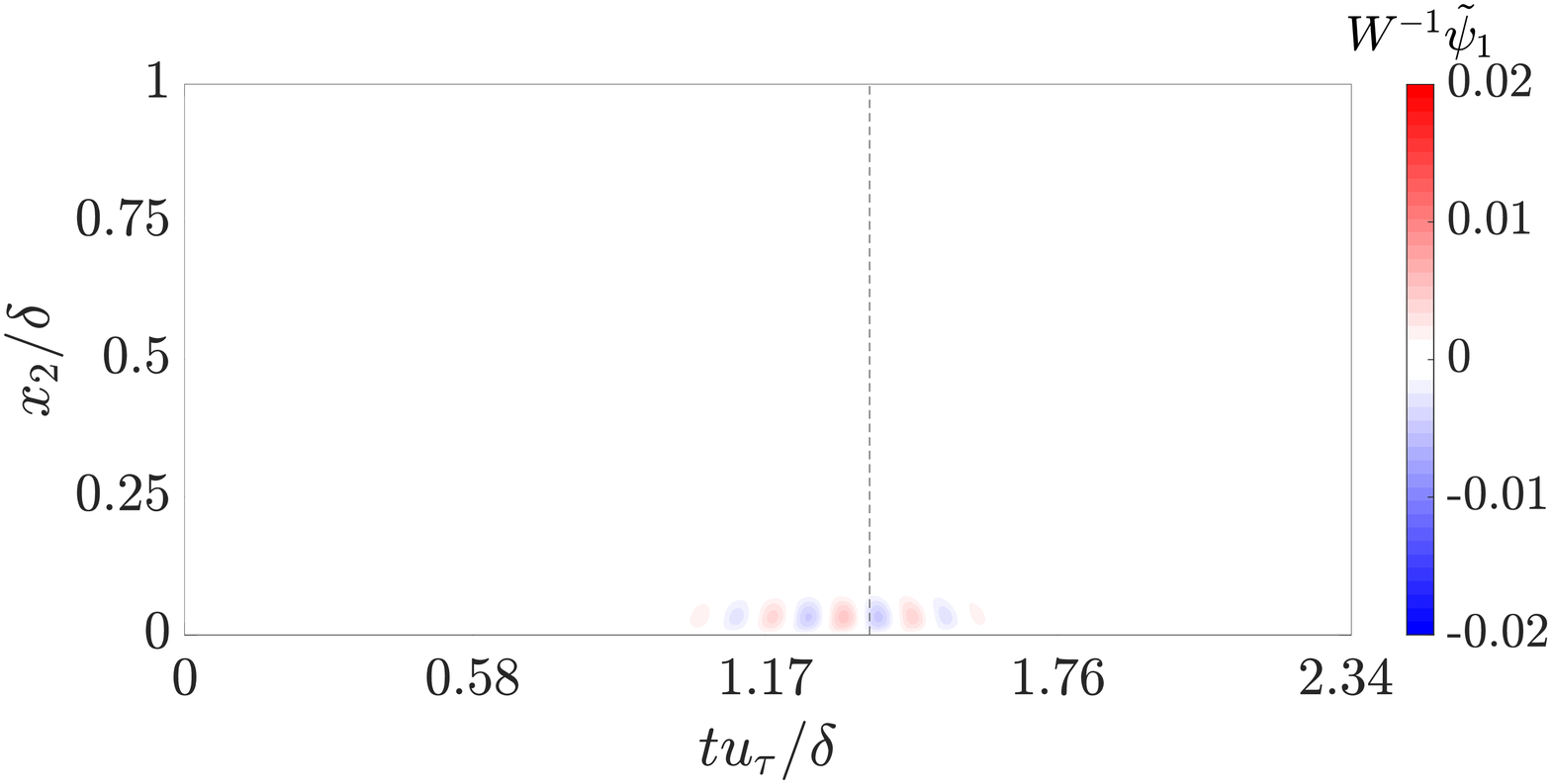}
    \caption{}
\end{subfigure}
\begin{subfigure}{0.45\textwidth}
    \includegraphics[width=\linewidth]{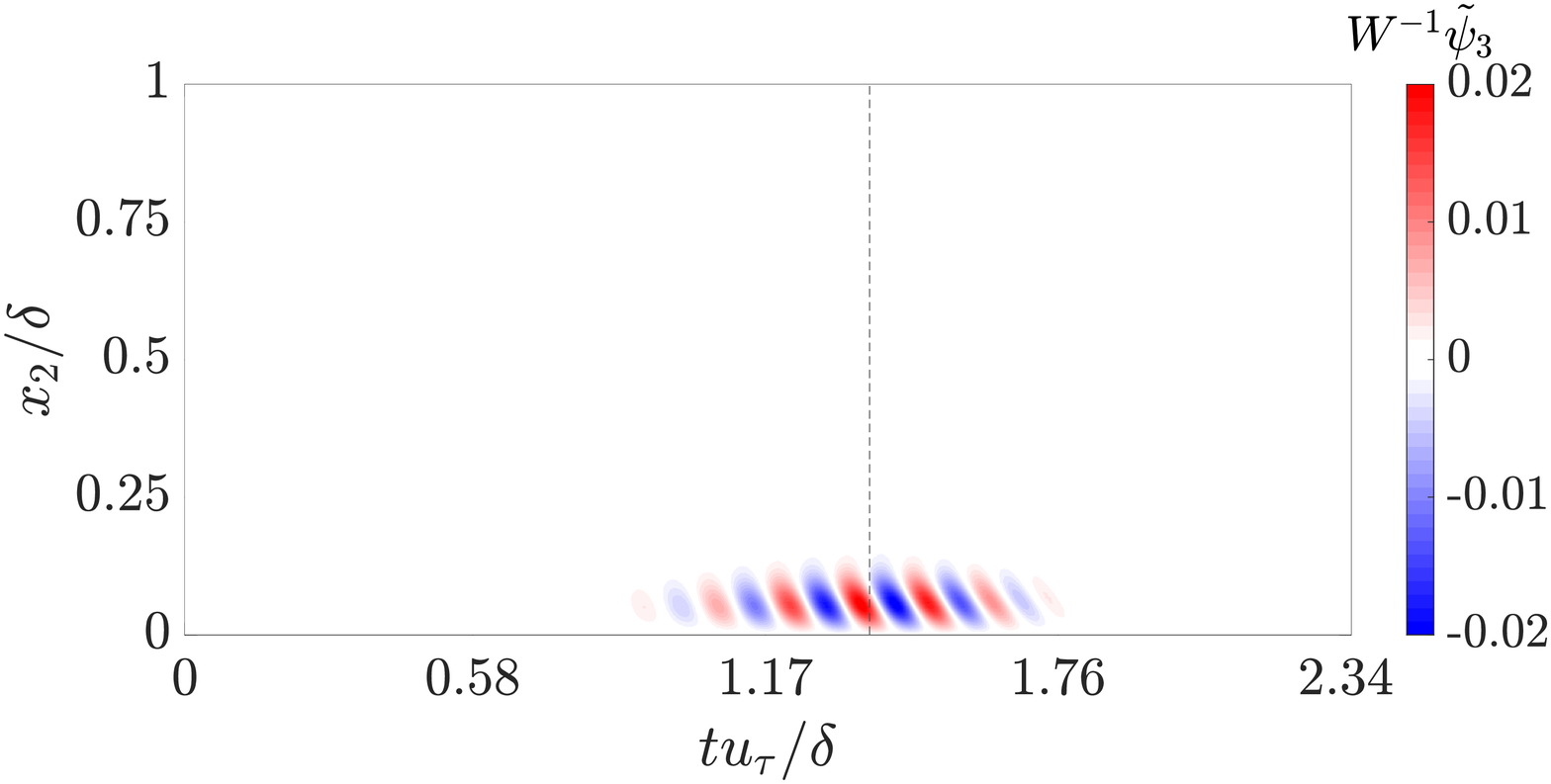}
    \caption{}
\end{subfigure}
\begin{subfigure}{0.45\textwidth}
    \includegraphics[width=\linewidth]{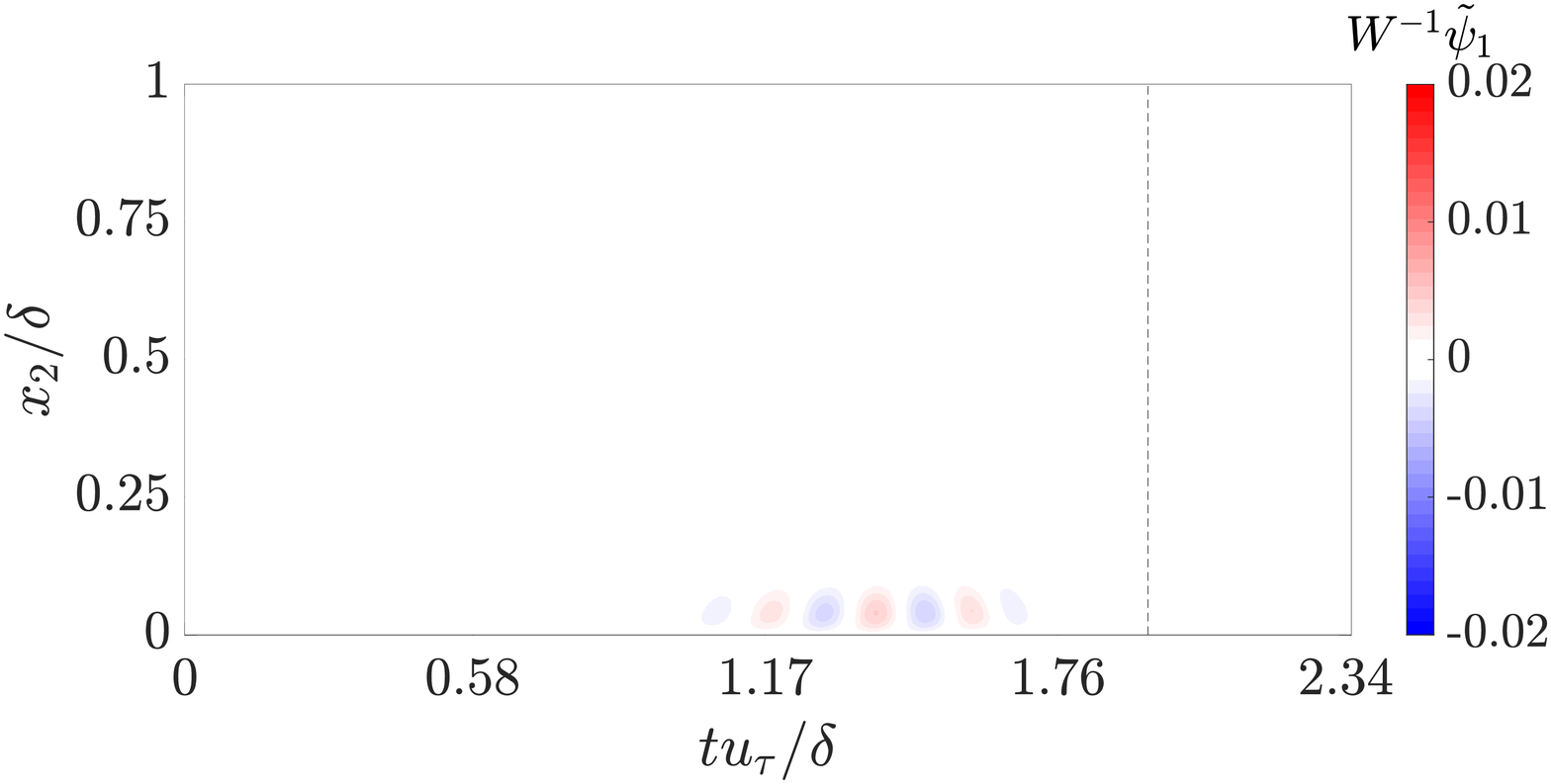}
    \caption{}
\end{subfigure}
\begin{subfigure}{0.45\textwidth}
    \includegraphics[width=\linewidth]{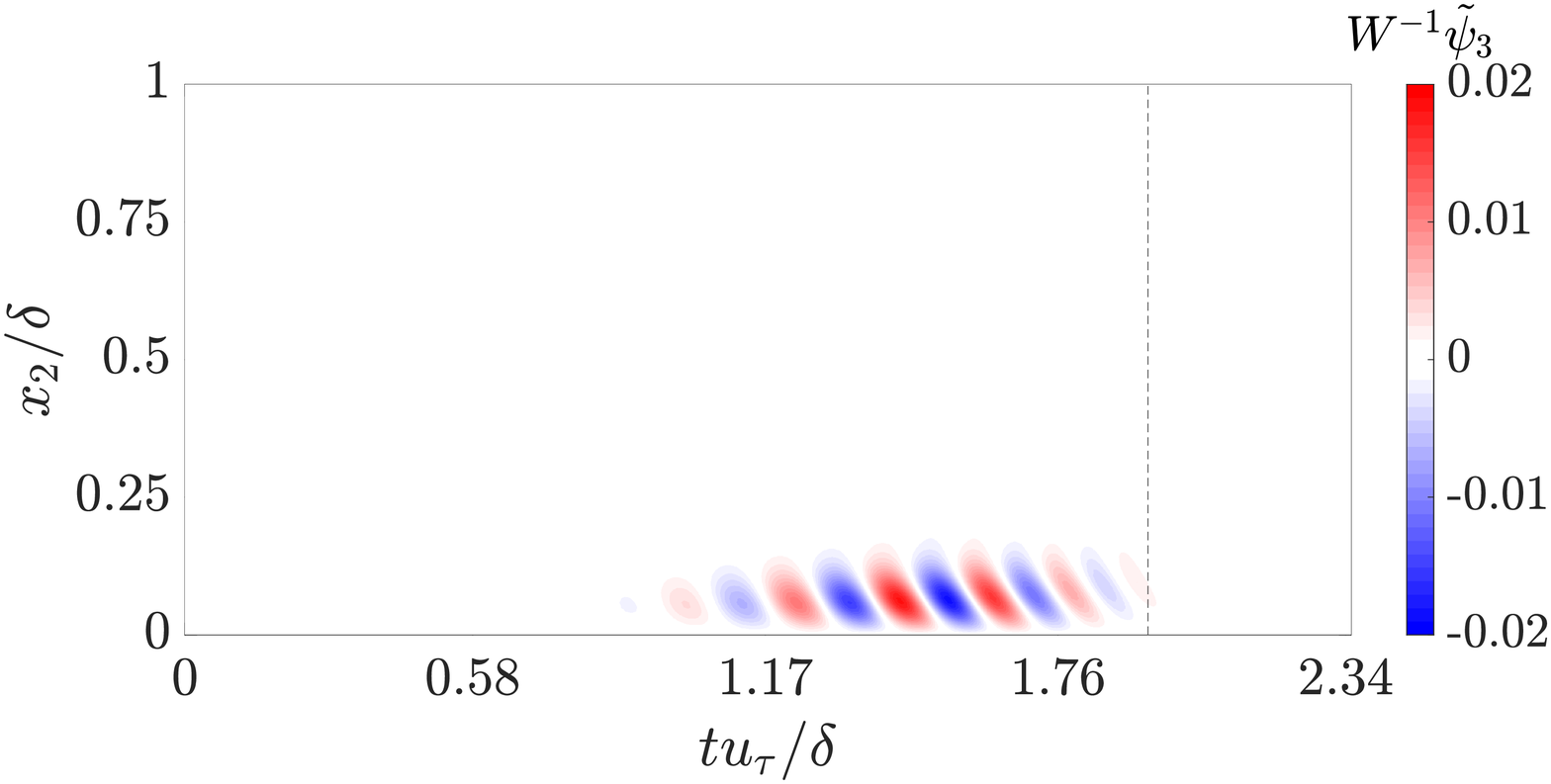}
    \caption{}
\end{subfigure}

\caption{Streamwise (left) and spanwise (right) velocity components of the resolvent output modes, for spatial scales of (a, b) $\lambda_1^+ = 1000$, $\lambda_3^+ = 100$, (c, d) $\lambda_1^+ = 189$, $\lambda_3^+ = 1890$, and (e, f) $\lambda_1^+ = 297$, $\lambda_3^+ = 2970$. The vertical dashed lines mark (a, b) $t\Omega = 0$, (c, d) $t\Omega=1.38$, and (e, f) $t\Omega=1.94$}
\label{fig:3DChannel}
\end{figure}

\begin{figure}
    \centering
    \begin{subfigure}{0.32\textwidth}
        \includegraphics[width=\linewidth]{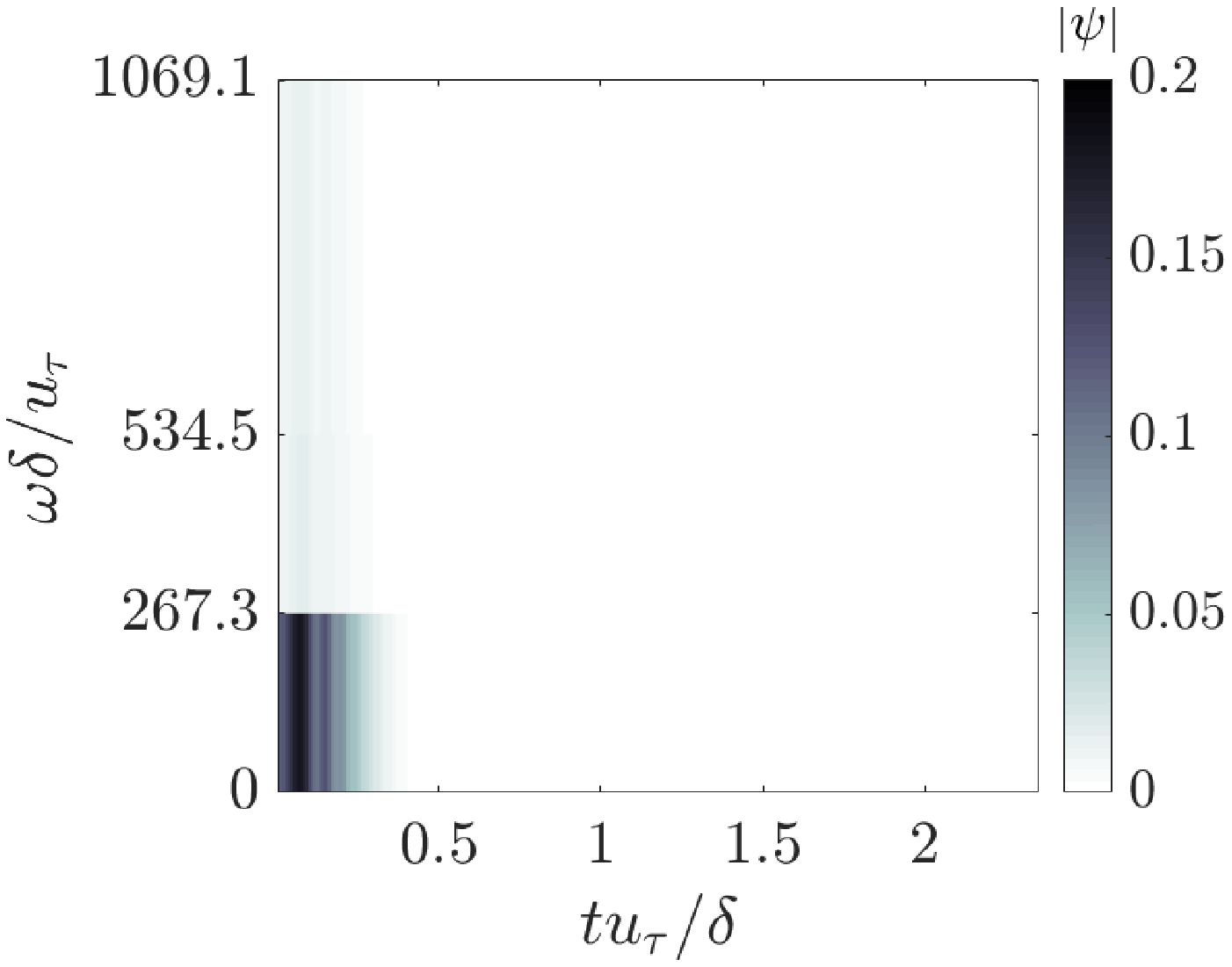}
        \caption{}
    \end{subfigure}
        \begin{subfigure}{0.32\textwidth}
        \includegraphics[width=\linewidth]{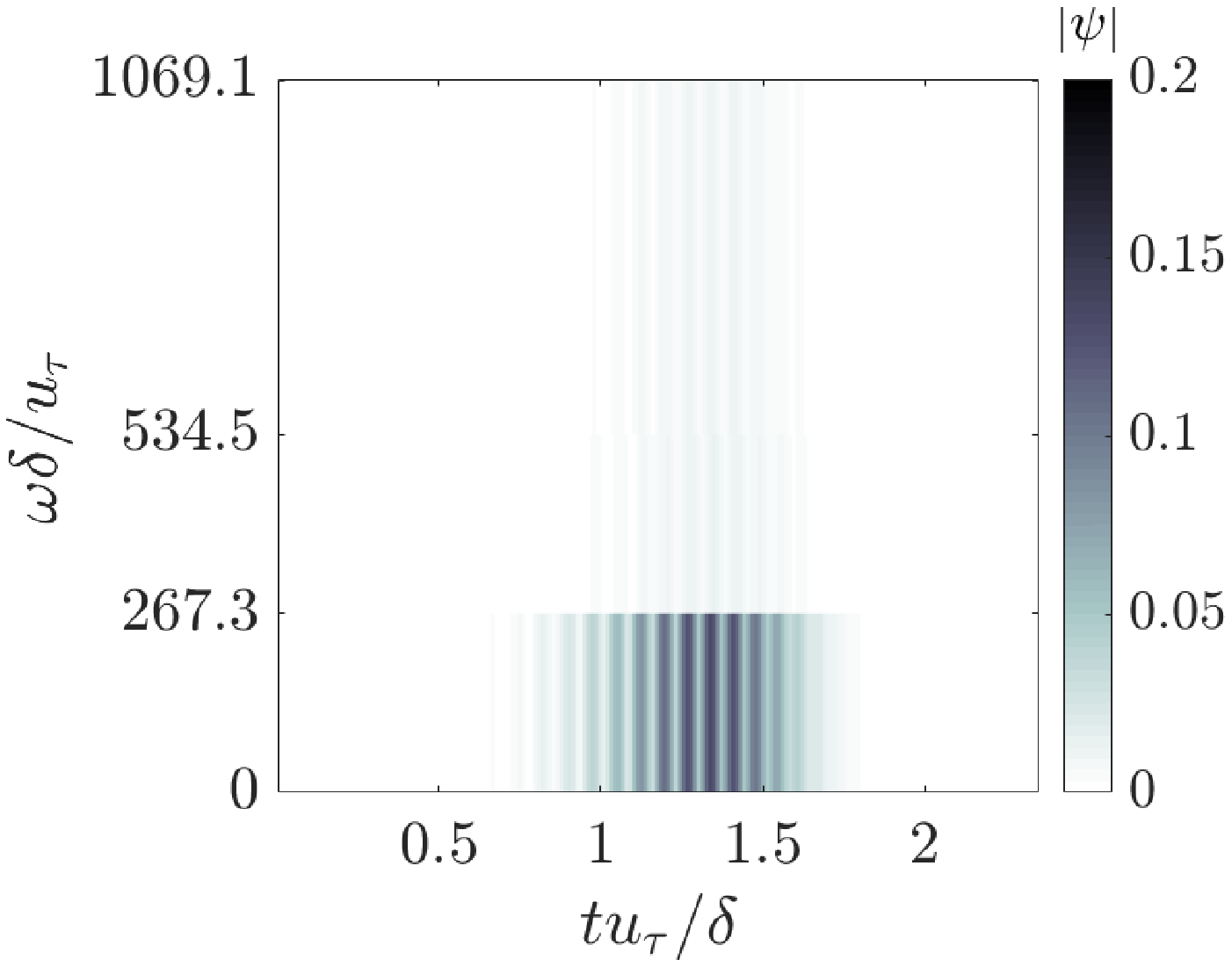}
        \caption{}
    \end{subfigure}
        \begin{subfigure}{0.32\textwidth}
        \includegraphics[width=\linewidth]{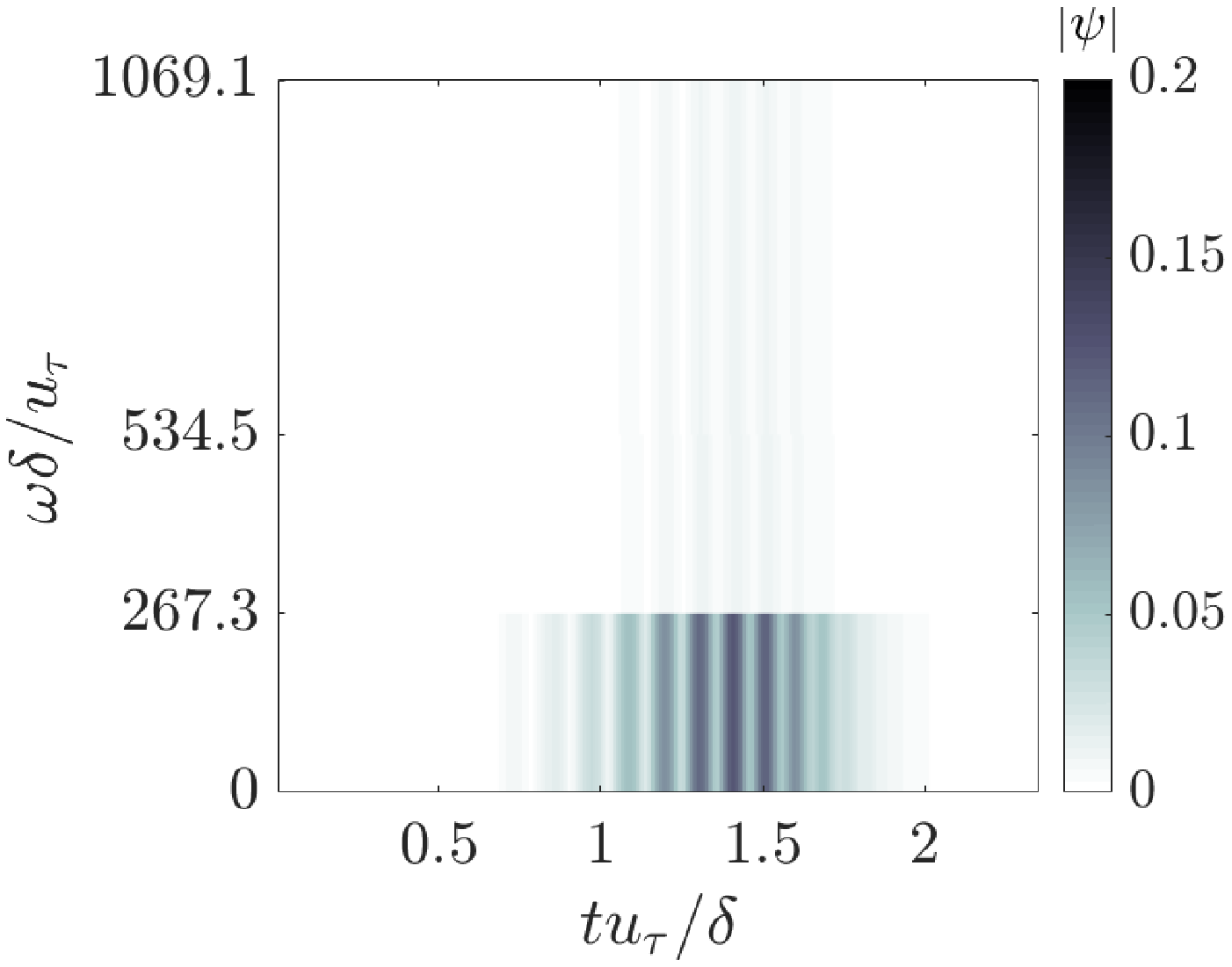}
        \caption{}
    \end{subfigure}

    \caption{Magnitude of the principal response mode in the frequency-time plane for (a) $\lambda_1^+ = 1000$, $\lambda_3^+ = 100$, (b) $\lambda_1^+ = 189$, $\lambda_3^+ = 1890$, and (c) $\lambda_1^+ = 297$, $\lambda_3^+ = 2970$.}
    \label{fig:3DScalogram}
\end{figure}

Regarding the spatial scales for the homogeneous directions, we choose $(\lambda_1^+,\, \lambda_3^+) = (1000,\,100)$ as well as ($\lambda_1^+,\, \lambda_3^+) = (189,\,1890)$ and $(\lambda_1^+,\, \lambda_3^+) = (297,\, 2970)$. Here, $(\cdot)^+$ indicates the wall scaling with respect to $u_{\tau,0}$, before the lateral pressure gradient is applied. The first combination of $\lambda_1^+$ and $\lambda_3^+$ is the same as in \S\ref{sec:app:channel} and corresponds to the spatial scales preferred by the near-wall streaks at $Re_\tau = 186$ prior to the lateral pressure gradient. The resolvent modes for these scales are shown in Fig.~\ref{fig:3DChannel}(a,b). The magnitude of the modes in frequency-time space is also plotted in Fig.~\ref{fig:3DScalogram}(a). The resolvent modes are temporally centered around $t = 0$ and exhibit a predominant streamwise component. The modes are located in a region  $x_2/\delta < 0.25$, which corresponds to $x_2^+ < 45$, i.e., the buffer region. Thus, at $t = 0$, the modes capture the highly energetic near-wall streaks. The subsequent temporal decay of these modes can be explained by the changing flow conditions, notably the growth of the spanwise wall-shear stress $\tau_3$, and consequently $u_\tau$ (see Figure \ref{fig:utau_angle}). Under these conditions, the spatial scales preferred by the near-wall streaks stretch as $u_\tau$ increases and the wall-shear stress tensor rotates toward the $x_3$ direction.

The second and third pairs of spatial scales are chosen so that the resolvent modes can capture the near-wall streaks under the new shear condition at times $t\Omega = 1.3$ and $t\Omega = 1.94$ respectively. To take into account the stronger mean shear in the spanwise direction, which increases proportionally to $dP/dx_3$ \cite{noneq}, the quantities in wall units must be scaled by a factor of $u_{\tau}(t)/u_{\tau,0}$. We also take into account the new orientation of the streaks by applying a rotation by the wall-shear stress angle $\gamma(t) = \tan^{-1}(\tau_3/\tau_1)$, where $\tau_i$ is the instantaneous wall-shear stress in the $x_i$ direction (see Fig. \ref{fig:utau_angle}). 
The new spatial scales are calculated by rotating a box of size $\lambda_1^+ = 100 (u_\tau(t)/u_{\tau, 0})$ and $\lambda_3^+ = 1000 (u_\tau(t)/u_{\tau, 0})$ with angle $\gamma(t)$ and finding the length scales aligned with the $x_1$ and $x_3$ axis. 
%The new spatial scales are then given by $\lambda_1^+(t) = 100 (u_\tau(t)/u_{\tau, 0})/\sin(\gamma(t))$, and  $\lambda_3^+ (t) = 1000 (u_\tau(t)/u_{\tau, 0})/\sin(\gamma(t))$.
The response mode for the second pair of spatial scales, $\lambda_1^+ = 189$ and $\lambda_3^+ = 1890$, correspond to $t\Omega = 1.3$ and are plotted in \ref{fig:3DChannel}(c,d). The frequency-time map of the modes is shown in Fig.~\ref{fig:3DScalogram}(b). Similar to the first case, the modes are centered around $t\Omega = 1.3$, indicating that the wavelet-based resolvent analysis is able to identify the nonequilibrium effects of the non-stationary flow. We note that the spanwise component of the response mode is much more dominant than the streamwise component, which reflects the new wall-shear angle $\gamma = 75.7^\circ$. We also observe that the modes are closer to the wall, which depict the reduction of the buffer layer as a result of the increase in $u_\tau$. 
Finally, for the third case, we choose $\lambda_1^+ = 297$ and $\lambda_3^+ = 2970$. We observe that these modes (Fig.~\ref{fig:3DChannel}e,f and Fig.~\ref{fig:3DScalogram}c) are not centered around $t\Omega = 1.94$. We speculate that this is due to the temporal boundary condition at $t\Omega = 2.34$. As the flow is not at a statistically-steady state at this time, a Neumann boundary condition may not be the most suitable boundary condition. The modes cannot grow beyond the boundary due to the boundary condition and are artificially damped near the end of the temporal domain.

\section{Summary} \label{sec:conclusion}
This work expands the resolvent analysis framework to non-stationary flow problems. The resolvent operator is traditionally constructed for flow quantities that are Fourier-transformed in the homogeneous spatial directions and in time. Such a resolvent operator cannot be used to study time-localized nonlinear forcing or a time-varying mean flow. Instead, we construct a wavelet-based resolvent operator, applying a wavelet transform in time while keeping the Fourier transform for the homogeneous spatial directions. This resolvent operator, provided we use an orthonormal wavelet basis, is equivalent to the Fourier-based resolvent analysis for statistically stationary flows. Even in stationary cases, wavelet-based resolvent analysis can be modified to explore the effects of transient forcing localized to time scales of interest, such as those characterizing the buffer or logarithmic layers. 

In the channel flow case, the wavelet-based resolvent modes are able to capture the transient growth expected for non-normal systems. We observe that the input modes precede the output modes, opening the possibility to study causality in turbulent flows using resolvent analysis. The wavelet-based resolvent analysis is notable in its ability to reflect the effects of a non-stationary mean flow. In the case of the Stokes oscillatory flow, the resolvent modes show increased sensitivity to forcing and perturbation amplification near the peaks of the streamwise root-mean-square velocity. The wavelet-based resolvent modes allow us to track the spatial and temporal location of the peak amplification alongside the varying mean flow. Finally, for the 3D channel flow, the resolvent modes are able to identify the effect of the varying flow conditions, mainly the increasing shear velocity and rotating wall shear stress, on the principal resolvent modes. We compute the resolvent modes using the length scales preferred by near-wall streaks for flow conditions at three different times. The resulting resolvent response modes peak around the chosen times, with the exception of the time close to the end of the temporal domain. The predominant velocity component for the resolvent modes also shifts from the streamwise component to the spanwise one, mirroring the reorientation of the mean flow. Wavelet resolvent modes reflect time-varying mean flow conditions and help locate energetic near-wall streaks in space and time, and identify their preferred spatial scales. This can shed light on the flow conditions that amplify these coherent structures. The cases considered in this work thus show the value of wavelet-based resolvent analysis as a new tool to study non-stationary turbulent flows.

\section*{Acknowledgments}
The authors acknowledge support from the Air Force Office of Scientific Research under grant number FA9550-22-1-0109.

\bibliography{references}

\end{document}